\documentclass[aps, prl,10pt,twocolumn,superscriptaddress,nofootinbib]{revtex4-1}
\usepackage{mathrsfs, amssymb, amsmath}  
\usepackage{epsfig, cancel}
\usepackage{nameref}
\usepackage{tabularx}
\usepackage{array}
\usepackage{latexsym}
\usepackage{array}
\usepackage{caption}
\usepackage{natbib, comment}
\usepackage{booktabs}
\usepackage{url}
\usepackage{eurosym}
\usepackage{dcolumn}
\usepackage{multirow}
\usepackage{color}
\usepackage{cancel}
\usepackage{soul}
\usepackage[normalem]{ulem}
\usepackage{amsfonts,amssymb,amsmath, txfonts}
\usepackage{graphicx,epsfig}
\usepackage{psfrag}
\usepackage{hyperref}
\hypersetup{colorlinks=true}
\usepackage{mathtools}
\usepackage{enumitem}
\usepackage{float}
\usepackage[dvipsnames]{xcolor}
\usepackage{xcolor}
\hypersetup{ linktoc=all,
    colorlinks, linkcolor={blue},
    citecolor={darkred}, urlcolor={darkgreen}
}

\usepackage{mathrsfs, amssymb, amsmath, mathtools} 
\usepackage{multirow}
\usepackage{amsfonts,amssymb,amsmath, txfonts}
\usepackage{graphicx,epsfig}
\usepackage{hyperref}
\usepackage{xcolor}
\usepackage{subcaption}
\usepackage{booktabs}
\graphicspath{{Images/}}
\newlength{\DefaultTabColSep}
\definecolor{rosy}{RGB}{230,235,252}
\definecolor{myframetitle}{RGB}{90,89,170}
\definecolor{myblocktitle}{RGB}{140,185,249}
\definecolor{mytitle}{RGB}{10,80,26}

\definecolor{darkgreen}{RGB}{27,130,45}
\definecolor{darkblue}{rgb}{0,0,0.3}
\definecolor{darkred}{rgb}{0.7,0,0}

\definecolor{light gray}{RGB}{220,220,220}
\definecolor{dark purple}{RGB}{108,0,217}
\definecolor{pink}{RGB}{190,20,100}
\definecolor{orang}{RGB}{193,63,0}
\definecolor{green}{RGB}{11,98,17}
\definecolor{darkpink}{RGB}{153,0,76}
\definecolor{bluegreen}{RGB}{0,102,102}
\definecolor{greenlagan}{RGB}{0,102,0}
\definecolor{redgreen}{RGB}{102,102,0}
\definecolor{Redgreen}{RGB}{153,76,0}
\definecolor{vividviolet}{rgb}{0.62, 0.0, 1.0}
\definecolor{amaranth}{rgb}{0.9, 0.17, 0.31}
\definecolor{palatinateblue}{rgb}{0.15, 0.23, 0.89}
\definecolor{brightpink}{rgb}{1.0, 0.0, 0.5}
\definecolor{cornflowerblue}{rgb}{0.39, 0.58, 0.93}
\definecolor{deepcarminepink}{rgb}{0.94, 0.19, 0.22}
\definecolor{radicalred}{rgb}{1.0, 0.21, 0.37}

\def\H0{{\text{H}\hspace*{-2.05mm}\text{H} 0\hspace*{-1.35mm}0\ }}

\def\be{\begin{equation}}
\def\ee{\end{equation}}
\def\beq{\begin{equation}}
\def\eeq{\end{equation}}
\def\bea{\begin{eqnarray}}
\def\eea{\end{eqnarray}}

\begin{document}

\title{Tracking High-Impact Researchers Through Highly Cited Tails}

\author{Hirushan Sajindra}
\email{hirushansajindra.buthpitilekamalage@research.atu.ie}
\affiliation{Atlantic Technological University, Ash Lane, Sligo F91 YW50, Ireland}
\author{Eoin \'O Colg\'ain}
\email{eoin.ocolgain@atu.ie}
\affiliation{Atlantic Technological University, Ash Lane, Sligo F91 YW50, Ireland}

\begin{abstract}
It is well recognised that field weighted citations are approximately lognormally distributed. 
This skewness reflects competition for attention, whereby most publications are gradually forgotten and only a minority continue to be recalled years later. 
Exploiting this observation, we construct a researcher-level indicator based on authorship of top 10\% cited papers in All Science Journal Classification (ASJC) bins and apply it to publications affiliated with Irish institutions over a typical funding cycle. Restricting attention largely to original research papers, we find that researchers authoring strictly more than one ASJC top-decile paper in a given year exhibit persistent participation in the highly cited tail of the literature in subsequent years, thus the simple track record indicator possesses predictive value. In contrast, the median researcher in cohorts of Research Ireland and European Research Council awardees selected through qualitative peer review either authors no top-decile paper or less than the median researcher in our ranking. These results demonstrate that research excellence as judged by qualitative peer review and influence on the research literature identify substantially different cohorts of researchers. In the absence of evidence, funding bodies and universities may conflate these two. 
While research reputation may be enhanced through many channels, highly cited papers constitute the most widely observable manifestation of international scholarly influence. 
Our analysis indicates that literature influence may be improved through greater use of data in track record pre-screening prior to qualitative peer review.  
\end{abstract}

\maketitle

\section{Introduction}
Building on the DORA declaration \cite{DORA:2012} and Leiden manifesto \cite{leiden:2015}, the Coalition for Advancing Research Assessment (CoARA) agreement \cite{coara:2022} advocates research assessment to be based primarily on qualitative judgement through peer review supported by the responsible use of quantitative indicators. The main thrust of the research reform movement \cite{DORA:2012, leiden:2015, coara:2022} is to focus assessment on research quality and pivot away from simple journal based metrics, primarily the journal impact factor (JIF) \cite{Garfield:1972}, which was never designed for research assessment \cite{Garfield:2006}, and $h$-index \cite{Hirsch:2005}.  

Traditionally, peer review is conducted by experts in the field but it has well-recognised shortcomings. These include low inter-reviewer agreement and poor reproducibility \cite{Marsh:2007, Bornmann:2010, Mutz:2012}, difficulties predicting impact \cite{Park:2015, Siler:2015, Fang:2016, Li:2017, wolfram:2021}, a lack of precision required for reliable rankings \cite{Kaplan:2008}, conservatism against new ideas \cite{Boudreau:2016, Lane:2022, Azoulay:2025}, penalisation of interdisciplinary research \cite{Porter:1985}, reputational and institutional bias \cite{Lee:2013, Tomkins:2017, Hultgren:2024}, limited accountability with a lack of a systematic feedback loop \cite{GAO:2022}, and finally high cost with poor scalability \cite{Aczel:2021, Pollitt:2026}. See \cite{Tartari:2022} for a review. In principle, data-based indicators can address a number of these shortcomings.  

Although field weighted citations cannot establish research quality \cite{Thelwall:2023}, they provide one with an internationally comparable record of academic uptake. Thus, the present work does not assume that citation impact and research quality are equivalent. Rather, it examines the extent to which influence on the research literature identifies the same researchers as qualitative peer review processes. The motivation is that society has become steadily more data-driven and one can attempt to apply the observation that field weighted citations are skewed \cite{Shockley:1957, Stringer:2008, Radicchi:2008, Evans:2012, Waltman:2012, Thelwall:2014, Golosovsky:2017} to both \textit{ex ante} research assessment for funding allocation and \textit{ex post} research evaluation to establish if academic/scientific impact was achieved. Peer review at the project level regularly handles the former but not the latter. Consequently, there is relatively little systematic feedback regarding the extent to which peer review based funding decisions predict subsequent literature influence.  

The basic idea of this paper is to identify researchers with original papers authored in the last funding cycle that are consistently in the tails of field weighted citation distributions at the national level. To ensure that most of the research is conducted in the jurisdiction of the funding body, we restrict our attention to papers with either a first or corresponding author from the jurisdiction. We apply our methods to the Irish context, where funding bodies favour assessment by review panels with limited recourse to quantitative indicators. 

A central question is the extent of overlap between researchers identified by our bibliometric indicator and researchers selected through qualitative peer review, both through generalist and expert panels. As noted by Li \cite{Li:2017c}, reviewers with field expertise better distinguish high and low-quality proposals but with a degree of bias in favour of their field. The concern is that reliance on generalists, while ostensibly at odds with CoARA's focus on research quality, makes it difficult to assess methodological feasibility, increasing the likelihood that funding decisions reflect the expertise available on the panel \cite{Porter:1985, Travis:1991, Materia:2015, Li:2017c, AbmaSchouten:2023, Oxley:2025}. This raises concerns about whether applicants compete on an equal footing. 

Although peer review may be the perceived gold standard, shortcomings remain, making it good practice to combine with quantitative indicators to correct bias, speed up and scale assessment in keeping with a data-driven world. A further concern is that generative AI can now assist in writing, polishing, tailoring and even reviewing proposals. By lowering the cost of proposal preparation, these tools are expected to increase application volume and place additional pressure on already stretched qualitative review systems. One response is to assess research impact through external databases combined with indicators, where the value is greater objectivity, reproducibility and scalability relative to peer review. Indicators can be specified \textit{ex ante}, before research begins, and applied consistently across all projects allowing comparison between projects. This limits the scope for success criteria to be tailored retrospectively through \textit{ex post} narratives and case studies, where assessment necessarily depends on subjective judgements by reviewers. While \textit{ex ante} peer review is a universal feature of research funding, systematic \textit{ex post} reassessment of \textit{individual} projects by independent reviewers is largely absent from contemporary research funding systems. 

Our work targets gaps in quantitative \textit{ex ante} assessment and \textit{ex post} evaluation in the Irish context. The goal is not to replace peer review with bibliometric indicators, but rather to provide a scalable mechanism for identifying researchers that have generated academic impact, thereby introducing an \textit{ex post} feedback loop that is largely absent from contemporary research funding systems. Our main output is a ranking of Irish researchers based on their capacity to author papers in the tails of field weighted citation distributions. Since papers greatly outnumber other research outputs, this is the most direct and visible channel to enhance international research reputation.

\section{Methodology}

\subsection{Philosophy}

We begin with overarching comments that motivate our approach. It is an empirical observation that field weighted citations for publications in a given year are skewed, being approximately lognormally distributed \cite{Shockley:1957, Stringer:2008, Radicchi:2008, Evans:2012, Waltman:2012, Thelwall:2014, Golosovsky:2017}. Moreover, it is also established that paper citations exhibit a universal life-cycle \cite{galiani:2017, Gou:2022} terminating in the paper being forgotten by the academic literature. ``Sleeping beauties", papers that go unnoticed for a long time before enjoying a rebirth of interest in terms of citations, provide exceptions but their rareness can be placed at $0.01\%$ of papers  \cite{vanRaan:2004, Ho:2017, Teixeira:2017}. The citation life-cycle provides a natural explanation for the skewed distribution. Most papers are gradually forgotten by the literature within a few years of publication, while a progressively smaller subset continue to be recalled, cited and reused. As a consequence, field weighted citation distributions become increasingly skewed with time. Nevertheless, despite early claims, e. g. \cite{Radicchi:2008}, citations cannot be described universally by either a lognormal or a power-law distribution since the tails continue to grow \cite{Golosovsky:2017}. 

Going further, there is little doubt that papers in the tails of field weighted citation distributions after 10-20 years have demonstrated significant influence on the literature through uptake of results. Provided one removes unoriginal research papers, primarily reviews, within this class one typically finds major discovery papers, method papers, data papers and foundational conceptual papers providing the motivation and scaffolding for later research. One could in principle mine this data to evaluate research and award funding, but the approach is retrospective and rewards researchers for impactful work done many years ago. A compromise is to mine the recent track record of academics over the last few years to see if they have authored papers that are on track to be in the tails of field weighted citations. While what constitutes being ``on track" is arbitrary, we choose papers in the top 10\% in field weighted citations in a given year. One could replace this with top 5\% or top 2\% to be more selective. We remind the reader again of the competition for attention and the rareness of sleeping beauties. While exceptions exist, papers that do not enter citation distribution tails early are unlikely to become highly influential at later times. 

\subsection{Methods}
We work nationally since public funding for research is typically confined to a jurisdiction. Although the national environment is generally less competitive than the international arena, international reputation is normally built upon national reputation. Alternatively put, researchers who fail to distinguish themselves within their own research system are unlikely to distinguish themselves internationally. 

Starting from SCOPUS metadata, we restrict the affiliation country to ``Ireland" and focus on publication types (article, conference paper, letter, note, data paper) that we are most confident are original contributions to the literature. Although conference papers may not be original research in the majority of fields, their importance to computer science warrants their inclusion. On the flip side, we removed books and book chapters, which are most important in Arts \& Humanities within social science. Compared to STEM subjects, books and book chapters are more prevalent in social sciences, but the ratio of books to articles is low and declining, and the general direction of travel is towards article publication \cite{Savage:2022}. We expect the removal of books and book chapters to have the greatest bearing on Arts \& Humanities with the All Science Journal Classification (ASJC) 12** and this caveat is worth bearing in mind. We provide a list of ASJC codes and subjects in the appendix. 

Having fixed the country and publication types, we restrict attention to publication years in the range 2022--2025 and export SCOPUS metadata including authors, author affiliations, document title, source (journal) title, citation count and correspondence address. All papers come with a unique identifier in the form of a SCOPUS URL, which we used for removing duplicate entries. SCOPUS provides a number of export formats, but only comma separated value (CSV) files preserve the mapping between authors and affiliations. In contrast, plain text (TXT) exports provide only a list of affiliations, preventing the attribution of affiliations to individual authors. SCOPUS allows exports up to 20,000 entries in one go, but we noted that there was a small percentage of duplicates and removed duplicate entries. 

One complication with the CSV export is that very large collaborations, primarily in physics and medicine, often contain hundreds or thousands of authors. When imported into Excel, the associated author and affiliation fields frequently exceed cell limits and are either truncated or overflow. To address this, large collaborations are processed manually. Interestingly, cell overflow in Excel provides an operational definition of a large collaboration. 

We digress momentarily to comment on large international collaborations. Research is, of course, an international endeavour. However, public funding agencies are generally expected not only to support scientific discovery but also to develop local research capacity and expertise. This consideration is particularly relevant for small countries such as Ireland. Large international collaborations, especially in physics and medicine, routinely produce publications with hundreds or even thousands of co-authors. In such settings, it can be difficult for a small national research group comprising a principal investigator, a postdoctoral researcher and a small number of PhD students to achieve visibility through manpower alone. Influence is more likely to arise when the collaboration adopts ideas, methods, software, instrumentation or analyses developed by the group. Consequently, smaller nations benefit most from research that can be conducted locally but nevertheless attracts international attention.

To identify research with a substantial local component, we focus on publications for which either the first author or corresponding author has an Irish affiliation. While this criterion is imperfect, we expect it to identify publications for which a significant fraction of the intellectual leadership, experimental work, analysis or project coordination was undertaken in Ireland. We therefore use first and corresponding author affiliation as a practical proxy for local research capacity and local expertise.

Having explained our motivation for the first/corresponding author restriction, we return to the SCOPUS CSV export, which we convert to Excel. For large collaborations that overflow the cells, we manually inspect the corresponding author and delete all entries without an Irish corresponding author. We also allow for an Irish first author if the author ordering is not alphabetical. In practice, this removes all but one or two of the large collaborations, which we carefully truncate to Excel cells so that the Irish authors and their affiliations are retained. Once this is done, all information is then in the correct columns and we resort to Python to process the file. All our files including the raw data from which our analysis begins,  \texttt{ireland\_4years\_data.xlsx}, which we exported from SCOPUS in early September 2026, can be downloaded from a Zenodo repository \footnote{\href{https://zenodo.org/records/23216889}{Ireland Researchers by Top 10\% Field Weighted Citations}}. 

Given the raw data, we remove all papers without an Irish first or corresponding author. Authors and affiliations are listed in the ``Authors with affiliations" column using the repeating pattern AUTHOR 1, AFFILIATION 1; AUTHOR 2, AFFILIATION 2; etc, where comma separates author and affiliation and distinct authors are separated through semi-colons. Thus, the first author's affiliation terminates at the first semi-colon, and separating the affiliation further on commas, our code checks if the last part is exactly ``Ireland". The corresponding author syntax for the SCOPUS export forms a repeating triplet pattern, AUTHOR; AFFILIATION; email: EMAIL, which repeats if there is more than one corresponding author. As before, we parse the AFFILIATION into parts using commas and check if the final part is exactly ``Ireland". In addition, we run checks on both the first author and corresponding author to ensure data quality, exporting all syntax violations including emails without @ symbols. 

Our SCOPUS exports have approximately 17,000 papers each year but restricting to an Irish first or corresponding author to ensure more of the research is done in Ireland reduces this to approximately 10,000 papers, which would seem to indicate that Ireland boosts paper production by 70\% through international collaboration even when large collaborations are mainly removed. From there we use the Elsevier SCOPUS ``Source title" export with journal ASJC codes \footnote{www.elsevier.com/products/scopus/content\#4-titles-on-scopus} to exactly match on the journal name and assign ASJC codes to each paper. Since journals typically span more than one ASJC code, we assign a score of 1 to each paper but divide by the number of ASJC codes $n$, so papers get a fractional score $1/n$ in each ASJC. We export all journal titles that do not match for inspection. The bulk of the non-matching papers are conference proceedings not in the Elsevier source title list, but a subset are non-matching journals that arise from differences between the source title file and the SCOPUS paper export. We correct for the latter through journal aliases that ensure matching. 

Ignoring the non-matching conference papers for the moment, we assign the citations and fractional score of each paper to the corresponding ASJC and retain only papers that appear in the upper decile or top 10\% of citations in each ASJC through a citation cutoff. Since each paper may be assigned to multiple ASJC codes through the publishing journal, a single paper contributes citations to multiple ASJC codes. Note, we include all papers at the citation cutoff, so in practice this inflates the number of retained papers beyond 10\% in each ASJC.  

To identify the number of papers to retain we divide the total number of papers in the ASJC by 10 and round to the nearest integer. The only exception to this practice is ASJC codes with 4 or less papers, which would get rounded down to zero, but for completeness we include a representative paper. This may introduce a small advantage for authors writing papers in obscure subfields, but it cannot greatly impact the output. Note, one could replace a top 10\% requirement by a top 5\% or top 2\% requirement. With a top 10\% requirement, one needs to think about ASJCs with 10 papers or less, but with more stringent requirements, one needs to think about ASJCs with 20 and 50 papers or less. There are many subfields where Ireland does not write 50 papers in a given year, so this would greatly reduce the scope if one projected the subjects out of the analysis. 

It is also important to note that publications are frequently assigned to multiple ASJCs. Consequently, selection occurs at the ASJC level rather than the paper level. In practice, retaining the top 10\% of papers within each ASJC results in approximately 2,500 retained ASJC-paper combinations from roughly 10,000 papers with an Irish first or corresponding author. The effective retention rate at the level of unique papers is therefore substantially greater than 10\%, and is closer to 25\%. The interpretation is that a paper on average is assigned to $2.5$ ASJC codes. The key point is that our indicator selects the 25\% most visible papers in terms of citations. This is not very selective and this makes our later results on funding awards all the more surprising. 

Having run the code once, we have a citation cutoff for each ASJC but still approximately 700 unmatched conference proceedings, largely from computer science. These unmatched conference papers affect our results in two ways. First, there is a small number of highly cited conference papers exceeding our citation cutoffs that should be included. Secondly, there is a large number of poorly cited conference papers that once assigned to fields inflate the volumes, thereby lowering the top 10\% threshold. To assign ASJC classifications to unmatched conference proceedings, we use Microsoft Copilot to suggest candidate ASJC codes based on conference titles. The expectation was that conferences overlapped with the computer science ASJC codes 17** and the recommendations are manually reviewed prior to inclusion in the source title alias file.

More precisely, we ensure all highly cited conference papers are assigned an ASJC. For the lowly cited papers, we prioritise conference proceedings with the greatest number of papers and iterate, running the code each time, until we are left with approximately 200 unmatched conference papers in conference proceedings with 2 or less papers. We employ the same strategy for each year of data we analyse. Note, given that we started with 10,000 odd original research papers with either an Irish first or corresponding author and only 200 odd lowly cited conference papers are omitted, approximately 98\% of the eligible publication set is assigned an ASJC classification.

Once we have the papers in the top 10\% in each ASJC, we add up the fractional scores for the paper, where the final score is capped at 1 by construction. A paper in a journal with $n$ ASJCs scores $m/n$ where $m \leq n$ is the number of ASJCs where the paper is the top 10\%. The \texttt{final\_top10percent\_papers\_by\_ASJC.xlsx} Excel file in the repository provides a breakdown of all papers in the top 10\% in each ASJC. We remove all overseas authors from the paper and only retain Irish researchers, their SCOPUS IDs and their affiliations. We also document the SCOPUS ID of the first and corresponding author of the paper irrespective of whether the researcher is based in Ireland or not. We use the latter information for later scoring. 

Once this is done, it is easy to produce a ranking of Irish authors each year. For each ASJC we assign the fractional score of the paper to all Irish authors and sum over all ASJCs, before totting up the scores for the authors. Note, each Irish author benefits equally from a top 10\% paper, where the goal is to measure participation in influential research rather than individual contribution. The results appear in \texttt{final\_author\_scores.xlsx}, where we rank by the score of the author. For the affiliations, we initially export all affiliations of the authors from \texttt{final\_top10percent\_papers\_by\_ASJC.xlsx} before matching them exactly to aliases for the largest institutions spanning both universities, hospitals, national laboratories, etc. Given that some of the TUs were only established in 2022, our aliases also map legacy institute of technology names. We also record the number of papers on which the author appears with an Irish first or corresponding author. This number is instructive as it provides a maximum score for the author if all papers appear in the top 10\% across all ASJC codes. We further document the percentage of the score of the author coming from papers where the author is a first and corresponding author. Authors with scores of zero in these two columns perform collaborative roles. Finally, we break down the score of each author by the ASJC ordered by the ASJC that contributes the most to the score. This helps to characterise the areas of expertise of the author. 

As a final comment, there is an inherent ambiguity in our method, but it is there in any method that matches a paper through its journal to an ASJC code. For SCOPUS' field weighted citation impact (FWCI) metric this ambiguity is handled by calculating an FWCI for each ASJC and then averaging. In our case, we assign the citations of the paper with a fractional score to all ASJCs of the journal. It should be borne in mind that some papers submitted to journals are truly multidisciplinary and may span all the ASJC subjects assigned to the journal. Nevertheless, authors make a choice when they submit to journals and they can choose a journal with a single ASJC over a journal with a broader focus across multiple ASJC codes.  

\subsection{Comments on Data Quality}

SCOPUS occasionally produces ``Author with affiliations" entries of the form AUTHOR 1; AUTHOR 2, AFFILIATION 2, \dots. This introduces ambiguity as AUTHOR 1 can be either an unaffiliated independent researcher or a researcher at AFFILIATION 2. For this reason, we flagged and reviewed all first authors without an affiliation manually and inserted the affiliation. It is our recommendation that SCOPUS removes this ambiguity in metadata by producing an affiliation for each author. 

In tandem, we flagged cases where the corresponding author syntax was violated. This most common violation coincided with a missing affiliation, which we corrected manually using the other fields, but it also included cases where the email was a URL (no @ symbol) and cases where the email address of the editor was confused as a corresponding author. This presumably stems from the use of AI in processing publications. By visually inspecting the publications we corrected the data. We also noted that SCOPUS occasionally truncates ``The United Kingdom of Great Britain and Northern Ireland" to ``Ireland" leading to affiliations such as ``Lancaster University, Ireland" and ``Oxford University, Ireland". This leads to impurities in our pipeline, but this is not a problem with the pipeline and simply an error for SCOPUS to resolve, especially since access to the SCOPUS database is sold.  

Data quality is also undermined by authors themselves. We employ a long list of aliases to map variants of university names. While the vast majority of authors follow official names, we identified all sorts of interesting variants, including ``Muster Technological University", ``Univercity of Limerick", ``University Colledge Dublin" and ``University Collage Dublin". Note, these mistakes occur at the level of the publication and not SCOPUS metadata. Although individually minor, such inconsistencies complicate institutional bibliometric analyses and suggest that greater attention to institutional branding is warranted. Ultimately, it is not too much that PhD level researchers can correctly spell the name of the host institution and induction may need to start with pedestrian matters.

\subsection{Summary}
We constructed a ranking of all researchers in Ireland active in the SCOPUS database based on the number of top 10\% papers in field weighted citations across approximately 330 ASJC subfields (see appendix). Given that papers can appear in multiple ASJCs, our methods retain approximately 25\% of papers, which is not very selective. The score of a researcher is limited by the number of papers they can co-author in a year with either an Irish first or corresponding author. One can of course improve this number through branching out into new fields, increasing collaboration, securing the funding to build a group and developing a deeper understanding of the wider literature. In contrast to SCOPUS' FWCI metric, we do not require an international normalisation, since the goal is not to identify internationally competitive researchers, simply the most competitive researchers locally in a funding jurisdiction. We enumerate a number of improvements over raw citation counts:

\begin{enumerate}[label=(\roman*), itemsep=0pt, topsep=2pt]
\item Introduction of first/corresponding author proxies makes research conducted in Ireland integral to study. 

\item Ranking accounts for differences in subject area citation practices. 

\item High publication volume cannot compensate for an absence of influential papers.

\item Non-original research articles as a route to accumulating citations are excluded. 

\item Comparison between junior and senior researchers is less sensitive to career length. 

\item Traditionally, citation counts are cumulative, but here our metric resets each year. 

\item Since sleeping beauty papers are rare \cite{vanRaan:2004, Ho:2017, Teixeira:2017}, very few influential papers are omitted. 

\item The influence of self and cartel citations in established fields is reduced relative to work in expanding fields and papers generating genuine interest. Self and cartel citations are expected to be a greater factor when citation volumes are small. As citation volumes accumulate over time, their relative influence diminishes, making highly cited tails increasingly difficult to sustain without broader uptake by the research community.

\end{enumerate}  

Collectively, these design choices produce a researcher-level indicator intended to capture persistent participation in influential research while remaining robust to differences in field citation practices, career length and publication volume.

\section{Trends in Data}
Before turning our attention to how funding bodies perform in the next section, in this section we outline insights that can be gleaned from the data.

\subsection{ASJC Counts}
We begin with an analysis of ASJC counts for papers that we have identified with either a first or corresponding author affiliated to Ireland. As explained, the rationale was to focus on research where one can be more confident that Ireland has made a significant contribution. Thus, the trends we illustrate here are expected to be a good indicator of research conducted in Ireland. Here we have a choice to rank by the number of papers that overlap with each ASJC code, but since each paper gets expanded across ASJC codes, it is better to rank by the total fractional score of all papers in each ASJC. The use of fractional counts ensures that multidisciplinary journals do not artificially inflate ASJC volumes. Although the publication set expands to multiple paper and ASJC assignments, the total fractional ASJC count is a better indicator of the number of papers.

\begin{figure*}[t]
\centering
\includegraphics[scale=0.5]{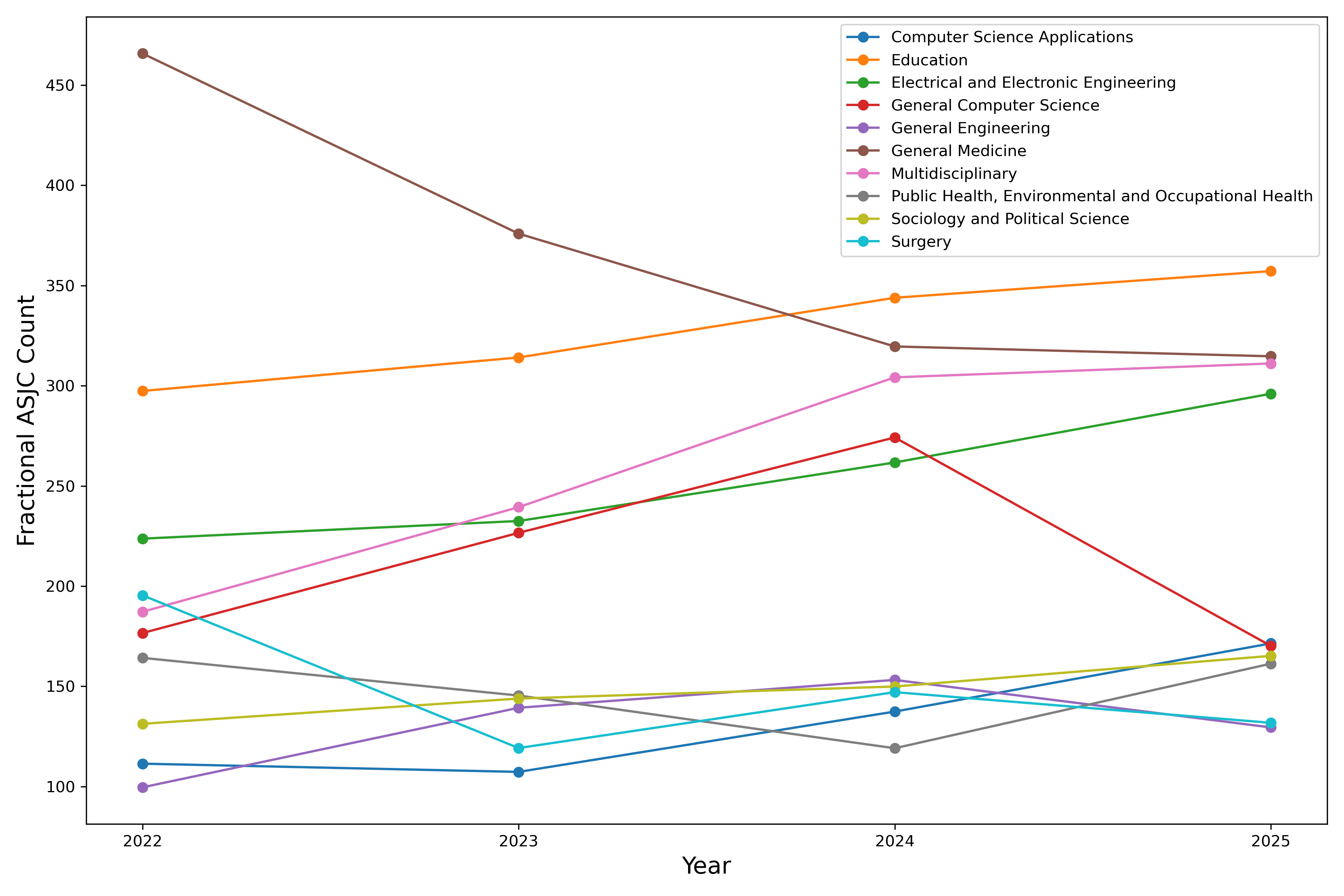}
\caption{Changes in fractional ASJC count of the ten subjects with the largest average fractional ASJC counts.}
\label{fig:asjc}
\end{figure*}

As an example, a paper with four ASJC codes contributes equally to all codes as a fractional count of 0.25, thereby ensuring that the total contribution of the paper across all ASJCs remains unity, and we sum up all the fractional scores in each ASJC code. Since the total fractional ASJC count of all ASJC subjects varies by year, we identify the subfields with the largest volumes by averaging over the four year period 2022--2025. We illustrate the most prominent subjects and their total fractional ASJC count each year in Fig. \ref{fig:asjc}.

It is also interesting to look at which subjects are increasing and decreasing in activity across the four year window. To do this, we demand that the ASJC appears in all four years and fit a line $y(x) = m x + c$ to the data to extract a best fit slope $m$ and error $\Delta m$, where $y$ is the total fractional ASJC score in each year and $x$ is the year. We then rank by the most statistically significant deviations from a zero slope through positive $m/\Delta m$ values to identify the growing areas and rank by the most negative $m/\Delta m$ values to identify the declining areas. Since $m/\Delta m$ is related to the Pearson correlation coefficient $r$ \footnote{For least squares fitting with $n$ points, we have $ \frac{m}{\Delta m} = \frac{r \sqrt{n-2}}{\sqrt{1-r^2}}$ \cite{DraperSmith:1998}.}, one could equivalently rank by $r$. The results are shown in Table \ref{tab:growing_asjcs} and Table \ref{tab:declining_asjcs}.

\begin{table}[t]
 
\centering 
\begin{tabular}{ccc} \hline\hline Subject & $m/\Delta m$ & Avg.\ Frac.\ ASJC\\
\hline 
General Agricultural and Biological Sciences & 16.2 & 14.6 \\  
Education & 9.7 & 328.2 \\ 
Sociology and Political Science & 9.0 & 147.6 \\
Pathology and Forensic Medicine & 7.9 & 18.6 \\ 
Cell Biology & 7.4 & 24.5 \\ 
Cardiology \& Cardiovascular Medicine & 6.9 & 102.9 \\ 
Anesthesiology and Pain Medicine & 6.2 & 42.2 \\  
Oncology (Nursing) & 6.1 & 14.0 \\ 
Computer Vision and Pattern Recognition & 6.1 & 23.9 \\
Community and Home Care & 5.9 & 5.1 \\ 
\hline\hline 
\end{tabular}
\caption{10 research areas exhibiting the most statistically significant increases in fractional ASJC count between 2022 and 2025. The trend significance is measured by the ratio of the fitted slope $m$ to its error $\Delta m$, $m/\Delta m$.}
\label{tab:growing_asjcs}
\end{table}

\begin{table}[t] 
\centering 
\begin{tabular}{ccc} \hline\hline Subject & $m/\Delta m$ & Avg.\ Frac.\ ASJC \\ \hline Instrumentation & -14.1 & 37.7 \\ Algebra and Number Theory & -12.0 & 15.9 \\ Management of Technology and Innovation & -9.1 & 28.2 \\ Geotechnical Engineering & -8.8 & 10.9 \\ History & -8.0 & 72.5 \\ Health (Social Science) & -6.8 & 59.1 \\ Biological Psychiatry & -5.9 & 11.4 \\ Cultural Studies & -5.0 & 51.0 \\ Pollution & -4.9 & 33.4 \\ Biotechnology & -4.6 & 34.9 \\ \hline\hline \end{tabular} 
\caption{10 research areas exhibiting the most statistically significant decreases in fractional ASJC count between 2022 and 2025. The trend significance is measured by the ratio of the fitted slope $m$ to its error $\Delta m$, $m/\Delta m$.}
\label{tab:declining_asjcs}
\end{table}

We can identify overlap between Fig. \ref{fig:asjc} and Table \ref{tab:growing_asjcs}. The subject areas of Education and Sociology \& Political Science exhibit large volumes and show a statistically significant increasing trend in Fig. \ref{fig:asjc}. In addition, from Fig. \ref{fig:asjc} General Medicine, Multidisciplinary, Electrical \& Electronic show noticeable decreasing and increasing trends, but fall just outside the top and bottom 10. It would be interesting to see if these increases and decreases are correlated with funding, but this takes us beyond the scope of the current paper.  

\subsection{Persistent Researchers}
As our next exercise, we turn to correlations between researcher scores across the years analysed. The goal is to understand how quickly one picks up predictive power and this brings us to our first main result.  In Table \ref{tab:score4plus} we present cohorts of researchers with a score $> 4$ on our ranking in any given year. As explained earlier, a score greater than 4 can be interpreted as a researcher who authors more than 4 whole papers in the top 10\% in field weighted citations in the year in question. From the \texttt{final\_author\_scores.xlsx} file in the repository it is clear that even when we restrict to Irish first/corresponding author papers, researchers in this cohort have the capacity to write a number of papers that are multiples of their top 10\% papers. Although it is beyond the scope of the current paper, it would be worth exploring what factors, e. g. funding, seniority, professional networks, research-group size, accumulated expertise, access to collaborators, etc, explain this capacity. 

Given that we have four years of data, we perform the analysis four times by using a single year's data to define the cohort and propagate the cohort forward and backwards in time. Throughout we define and match a researcher through their SCOPUS ID. We record lower and upper quartiles (P25, P75) and deciles (P10, P90), and the median for the score or number of top 10\% papers. We define a researcher to be active in a year if they author a paper in our raw data, essentially in the SCOPUS export metadata corrected for large collaborations, with an Irish affiliation. Note, papers cannot score unless they have an Irish first or corresponding author, so by this definition a researcher may be active but yet incapable of scoring. This may be a disadvantage if the goal was to analyse researchers, but our goal here is to analyse and compare cohorts, so this distinction is unimportant as long as we treat cohorts the same throughout. The introduction of active researchers allows researchers to enter and leave the jurisdiction, or simply stop publishing, without affecting the results. Active researchers who fail to rank are assigned the score zero. In all cases both the number of active researchers and our median score peak on the year used to construct the cohort. As can be seen from Table \ref{tab:score4plus}, in most years the lower decile is positive except for the 2022, which is the weakest year and we see this through a pronounced dropoff in scoring researchers relative to active researchers. 

\begin{table}[t] \centering \begin{tabular}{cccccccc} \hline\hline Year & Active & Scoring & P10 & P25 & Median & P75 & P90 \\
\hline\hline
\multicolumn{8}{l}{\textbf{2022 Score $ >4$ Cohort (29)}}\\ 
2022 & 29 & 29 & 4.47 & 4.50 & 5.00 & 5.50 & 6.17 \\
2023 & 27 & 24 & 0.30 & 1.23 & 2.00 & 3.58 & 4.24 \\
2024 & 27 & 23 & 0.00 & 1.00 & 1.50 & 2.79 & 4.75 \\
2025 & 25 & 17 & 0.00 & 0.00 & 1.00 & 3.00 & 4.00 \\ 
\hline
\multicolumn{8}{l}{\textbf{2023 Score $> 4$ Cohort (38)}}\\
2022 & 36 & 33 & 0.50 & 1.44 & 2.10 & 3.00 & 3.68 \\
2023 & 38 & 38 & 4.24 & 4.50 & 5.00 & 5.98 & 7.85 \\
2024 & 38 & 34 & 0.23 & 1.17 & 3.00 & 3.90 & 6.56 \\
2025 & 36 & 34 & 0.29 & 1.00 & 3.00 & 4.08 & 6.04 \\ 
\hline
\multicolumn{8}{l}{\textbf{2024 Score $> 4$ Cohort (25)}}\\
2022 & 24 & 21 & 0.15 & 0.98 & 2.17 & 3.19 & 5.40 \\
2023 & 25 & 23 & 0.90 & 1.75 & 3.67 & 5.00 & 7.87 \\
2024 & 25 & 25 & 4.53 & 4.83 & 5.00 & 6.20 & 9.87 \\
2025 & 25 & 24 & 1.00 & 1.67 & 3.50 & 5.92 & 7.93 \\
\hline 
\multicolumn{8}{l}{\textbf{2025 Score $> 4$ Cohort (26)}}\\
2022 & 23 & 20 & 0.07 & 1.00 & 2.08 & 2.92 & 3.55 \\
2023 & 25 & 24 & 0.80 & 1.67 & 2.70 & 4.67 & 6.90 \\
2024 & 25 & 23 & 1.00 & 2.00 & 3.25 & 6.00 & 9.87 \\
2025 & 26 & 26 & 4.38 & 4.67 & 5.00 & 6.00 & 7.92 \\
\hline
\hline
\end{tabular}
 \caption{Performance in window 2022--2025 of researchers with a score of 4 or greater in a given year in the window with the median or 50$^{\rm th}$ percentile (P50) peaking on the year sampled. Active researchers are those producing at least one paper with an Irish affiliation in the given year. Scoring researchers are active researchers with a non-zero top 10\% score and non-scoring active researchers are assigned the score zero. Percentiles are calculated using only active researchers.} \label{tab:score4plus}\end{table}

The main takeaway from the exercise is that the typical or median high scoring researcher in any given year continues to author top 10\% papers up to three years before and after the year used to define the cohort. This shows that authoring papers in the tails of field weighted citation distributions is a persistent researcher characteristic that allows one to guarantee that the typical researcher generates impact a few years later even based on a single year's data. Interestingly, the persistence of these researchers has both a predictive and retrospective property, whereby to be in a cohort of score $>4$ researchers in 2025, the typical researcher in this cohort writes approximately two top 10\% papers three years earlier in 2022. 

\begin{table}[t] \centering \begin{tabular}{cccccccc} \hline\hline Year & Active & Scoring & P10 & P25 & Median & P75 & P90 \\
\hline\hline
\multicolumn{8}{l}{\textbf{2022 $3 < \mathrm{\bf Score} \leq 4$ Cohort (33)}}\\ 
2022 & 33 & 33 & 3.21 & 3.33 & 3.67 & 4.00 & 4.00 \\
2023 & 32 & 21 & 0.00 & 0.00 & 1.00 & 1.81 & 4.63 \\
2024 & 28 & 22 & 0.00 & 0.44 & 1.00 & 2.26 & 3.62 \\
2025 & 28 & 17 & 0.00 & 0.00 & 0.50 & 1.00 & 1.99 \\
\hline
\multicolumn{8}{l}{\textbf{2022 $2 < \mathrm{\bf Score} \leq 3$ Cohort (110)}}\\
2022 & 110 & 110 & 2.25 & 2.50 & 2.74 & 3.00 & 3.00 \\
2023 & 105 & 73 & 0.00 & 0.00 & 1.00 & 2.50 & 4.12 \\
2024 & 98 & 65 & 0.00 & 0.00 & 1.00 & 2.00 & 3.44 \\
2025 & 95 & 64 & 0.00 & 0.00 & 1.00 & 2.00 & 3.90 \\ 
\hline
\multicolumn{8}{l}{\textbf{2022 $1 < \mathrm{\bf Score} \leq 2$ Cohort (388)}}\\
2022 & 388 & 388 & 1.25 & 1.50 & 1.75 & 2.00 & 2.00 \\
2023 & 335 & 195 & 0.00 & 0.00 & 0.50 & 1.08 & 2.30 \\
2024 & 321 & 171 & 0.00 & 0.00 & 0.33 & 1.00 & 2.00 \\
2025 & 308 & 160 & 0.00 & 0.00 & 0.25 & 1.00 & 2.00 \\
\bottomrule
\hline 
\multicolumn{8}{l}{\textbf{2022 Score $= 1$ Cohort (1999)}}\\
2022 & 1999 & 1999 & 1.00 & 1.00 & 1.00 & 1.00 & 1.00 \\
2023 & 1072 & 393 & 0.00 & 0.00 & 0.00 & 1.00 & 1.25 \\
2024 & 992 & 335 & 0.00 & 0.00 & 0.00 & 0.80 & 1.00 \\
2025 & 929 & 296 & 0.00 & 0.00 & 0.00 & 0.67 & 1.00 \\
\hline
\hline
\end{tabular}
 \caption{Predictive value of a single year of bibliometric performance. Researchers are grouped by their 2022 score and followed through 2025. A score of one has little predictive value, yielding median future scores of zero. In contrast, scores strictly greater than 1 are associated with persistent participation in the highly cited tail of the literature over multiple subsequent years.} \label{tab:prediction}\end{table}

Despite showing that the correlation works both forward and backwards in time, for funding it is most relevant to focus on predictions. In Table \ref{tab:prediction} using 2022 as the reference year, we define cohorts of researchers in different scoring brackets, where our goal is to identify a threshold where the typical (median) researcher no longer authors a top 10\% paper in a subsequent year. Once again the median decreases from 2022 but remains non-zero provided we restrict our attention to researchers with the capacity to write strictly more than one top 10\% paper a year. Combined with the score $>4$ cohort, this gives one a pool of $29+33+110+388 = 560$ researchers in 2022, which in this dataset, identifies a cohort whose median researcher continues to author a non-zero number of top 10\% papers in subsequent years. Further restricting to $29+33+110 = 172$ researchers with scores $>2$, the typical researcher continues to author approximately one top $10 \%$ paper in subsequent years. The likely interpretation of these results is that while one may write a single top 10\% paper through temporary circumstances, exceptional collaboration opportunities, or stochastic factors, writing more than one paper requires a persistent capacity to generate highly cited research. Thus, the main takeaway from the analysis is that a single top 10\% paper has little predictive value, whereas repeated appearances in the highly cited tail are associated with sustained future participation in the highly cited literature.

It should be stressed that Table \ref{tab:score4plus} and Table \ref{tab:prediction} are persistence statements based on only a single year's data. If one was going to award funding based on bibliometric data supporting qualitative peer review, it is better to focus on researchers who appear more than 1 year in the ranking. In Table \ref{tab:persistence} we identify SCOPUS IDs of researchers who appear 1, 2 and 3 years in our rankings in the three year window 2022--2024. We then track the cohort over the four year window 2022--2025 where the first 3 years are retrospective and the final year constitutes a prediction. 

There are two main observations from Table \ref{tab:persistence}. First, cohorts of reappearing researchers demonstrate greater retrospective performance over almost all percentiles, which is not so surprising given earlier results. The second is that the percentage of scoring researchers to active researchers increases as we demand greater persistence. As is clear from the 2025 results, as we increase from appearing 1 year, 2 years and 3 years in the 2022--2024 window, the percentage of scoring researchers in 2025 increases $1710/4814 \approx 36\%$ to $760/1589 \approx 48\%$ and $260/393 \approx 66 \%$. Thus, reappearance in the ranking acts as a simple but powerful filter, and requiring appearance in all three years nearly doubles the probability that an active researcher records a non-zero score in 2025 compared with requiring appearance in only one year. We note that the median provides a much cruder measure than the proportion of scoring researchers and only becomes non-zero when researchers reappear in all three years. Nevertheless, there is a clear lesson. Even without imposing a threshold, simply by demanding reappearance in the ranking, one can strengthen the performance of a cohort.     

\begin{table}[t] \centering \begin{tabular}{cccccccc} \hline\hline Year & Active & Scoring & P10 & P25 & Median & P75 & P90 \\
\hline\hline
\multicolumn{8}{l}{\textbf{1 from 3 years (8996)}}\\ 
2022 & 6236 & 3874 & 0.00 & 0.00 & 0.50 & 1.00 & 1.00 \\
2023 & 6187 & 3778 & 0.00 & 0.00 & 0.50 & 1.00 & 1.08 \\
2024 & 6161 & 3724 & 0.00 & 0.00 & 0.50 & 1.00 & 1.00 \\
2025 & 4814 & 1710 & 0.00 & 0.00 & 0.00 & 0.75 & 1.00 \\
\hline
\multicolumn{8}{l}{\textbf{2 from 3 years (1950)}}\\
2022 & 1793 & 1394 & 0.00 & 0.25 & 1.00 & 1.00 & 2.00 \\
2023 & 1862 & 1527 & 0.00 & 0.33 & 1.00 & 1.05 & 2.00 \\
2024 & 1808 & 1409 & 0.00 & 0.25 & 1.00 & 1.00 & 2.00 \\
2025 & 1589 & 760 & 0.00 & 0.00 & 0.00 & 1.00 & 2.00 \\
\hline
\multicolumn{8}{l}{\textbf{3 from 3 years (430)}}\\
2022 & 430 & 430 & 0.50 & 1.00 & 1.00 & 2.00 & 3.00 \\
2023 & 430 & 430 & 0.50 & 1.00 & 1.00 & 2.00 & 3.50 \\
2024 & 430 & 430 & 0.50 & 1.00 & 1.00 & 2.00 & 3.00 \\
2025 & 393 & 260 & 0.00 & 0.00 & 0.80 & 1.67 & 3.09 \\
\hline
\hline
\end{tabular}
 \caption{Distribution of scores for cohorts of researchers appearing multiple years in the ranking in the 2022--2024 window, where 2025 results represent a prediction.} \label{tab:persistence}\end{table}

\subsection{Leadership}
In this subsection we investigate the correlation between leadership and researcher score. Any peer-reviewed publication requires a first and corresponding author, with single-author papers representing an extreme case in which one individual fulfills both roles. These positions are therefore naturally associated with leadership and responsibility for the work. Where publications have multiple authors, we are conscious that authorship conventions vary across disciplines. In a small percentage of papers, estimated at less than 4\% across all fields and declining in 2012 \cite{Waltman:2012b}, authors are listed alphabetically. In such cases, first authorship conveys no information regarding contribution. However, for the overwhelming majority of publications, author order reflects contribution, allowing first authorship to be interpreted as an indicator of intellectual leadership. 

The role of the corresponding author is more nuanced. By definition, the corresponding author is responsible for communication with the journal. It has been demonstrated that corresponding author usually coincides with first author, but in recent years there is a trend towards middle and last authors \cite{ChinchillaRodriguez:2024}. While corresponding authors are often assumed to occupy senior intellectual or supervisory roles within a project, this need not always be the case. Corresponding authorship may also reflect administrative responsibilities, institutional publishing arrangements, or other practical considerations. Unlike the estimate of alphabetical authorship provided by Waltman \cite{Waltman:2012b}, we are unaware of any study that quantifies the prevalence of such practices or places an upper bound on their frequency.

\begin{figure}
\centering
\includegraphics[scale=0.5]{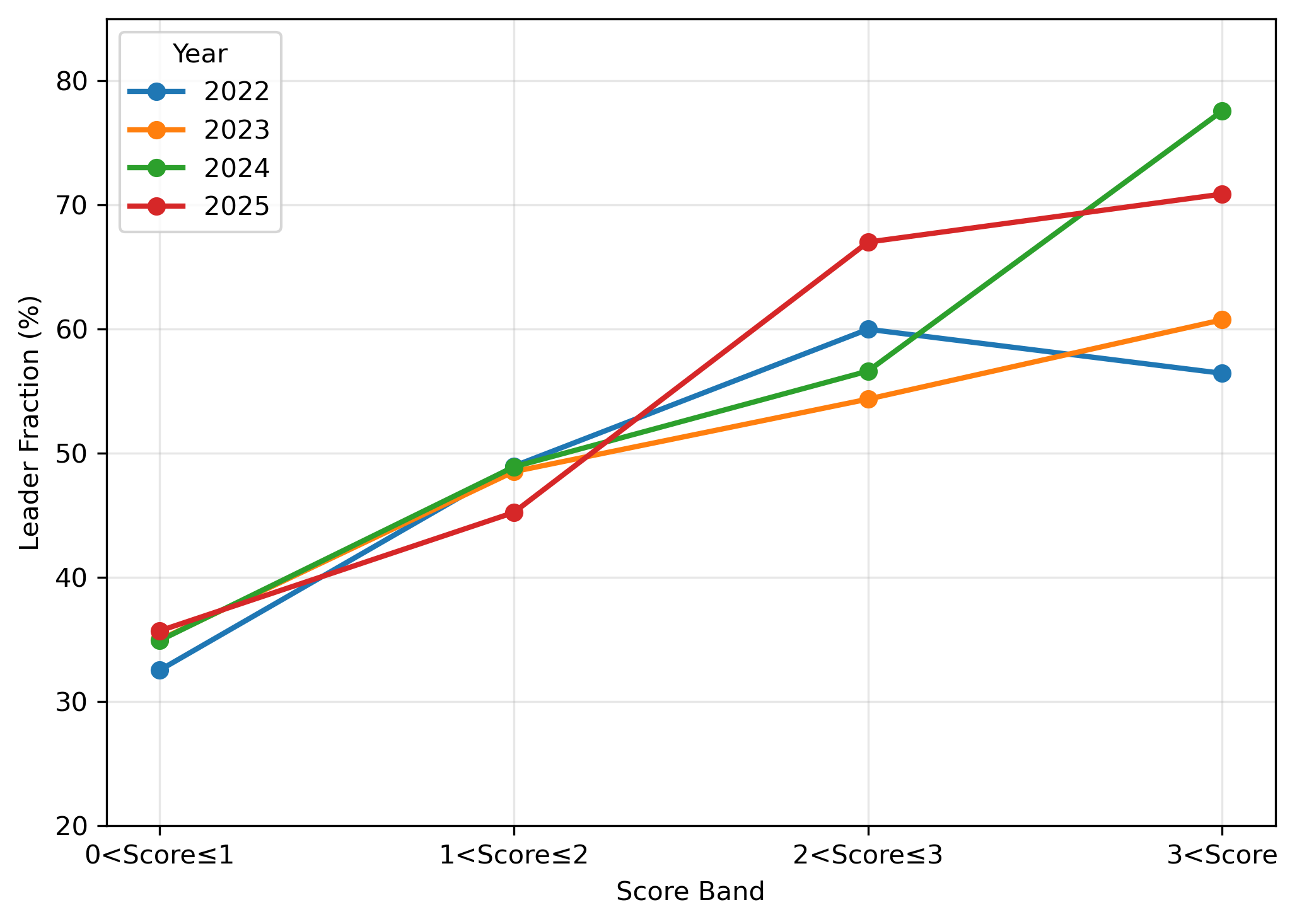}
\caption{Fraction of researchers classified as leaders as a function of researcher score. Leadership is defined by a non-zero first author or corresponding author score. The proportion of leaders increases systematically with score.}
\label{fig:leaders}
\end{figure}

Researchers are divided into leaders and non-leaders according to their contribution to top 10\% papers. Specifically, a researcher is classified as a leader if they possess a non-zero first author score or corresponding author score. Researchers with zero first author and zero corresponding author scores are classified as non-leaders. Figure~\ref{fig:leaders} shows the relationship between researcher score and leadership status. The proportion of leaders increases steadily with score, rising from approximately one third of researchers with scores below one to around two thirds of researchers with scores above three. This positive correlation is consistent with the expectation that researchers occupying first author or corresponding author roles are more likely to be involved in highly influential research outputs. At the same time, a substantial fraction of high scoring researchers remain classified as non-leaders, indicating that sustained participation in influential research is not confined to formal leadership authorship positions. 

\subsection{Identification of Emerging Researchers}
In each year, more than 3,700 researchers appear in the ranking. Attention naturally focuses on researchers at the top of the ranking, however the dataset also permits the identification of researchers whose performance is improving over time. To identify such researchers, we follow our earlier method for fractional ASJC counts and fit straight lines to annual researcher scores for individuals active in Ireland in at least three of the four years analysed. The requirement of three years of data is important, since a line fitted to only two points is exact and therefore provides no meaningful estimate of the uncertainty on the slope.

The first step of the algorithm is to extract the researcher's SCOPUS ID and annual score from each ranking. We then return to our raw SCOPUS exports, which we corrected for large collaborations, and identify researchers who were active in a particular year through a paper published with an Irish affiliation. This step allows us to distinguish researchers who were inactive in Ireland from researchers who remained active but did not contribute to any top 10\% paper. Consequently, if a researcher appears in the ranking for a given year, we assign their score. If the researcher is active in Ireland but absent from the ranking, we assign a score of zero. Finally, if the researcher is inactive in Ireland, we assign a value of NaN.

This process ensures that we have a list of four numbers or NaN for each SCOPUS ID and we restrict our attention to researchers with at least three non-missing observations. We then fit a line $y(x) = m x + c$, where $y$ denotes the annual score of the researcher and $x$ denotes the year, to extract the slope $m$ and its error $\Delta m$, before ranking researchers according to the largest positive $m/ \Delta m$ values. The quantity $m/\Delta m$ may be interpreted as a signal-to-noise ratio for the trend, with larger values providing stronger evidence that the underlying slope is positive. This identifies researchers with increasing annual scores, with priority given to slopes where the difference with zero slope characterising a horizontal is most statistically significant. 

\begin{figure*}
\centering
\includegraphics[scale=0.6]{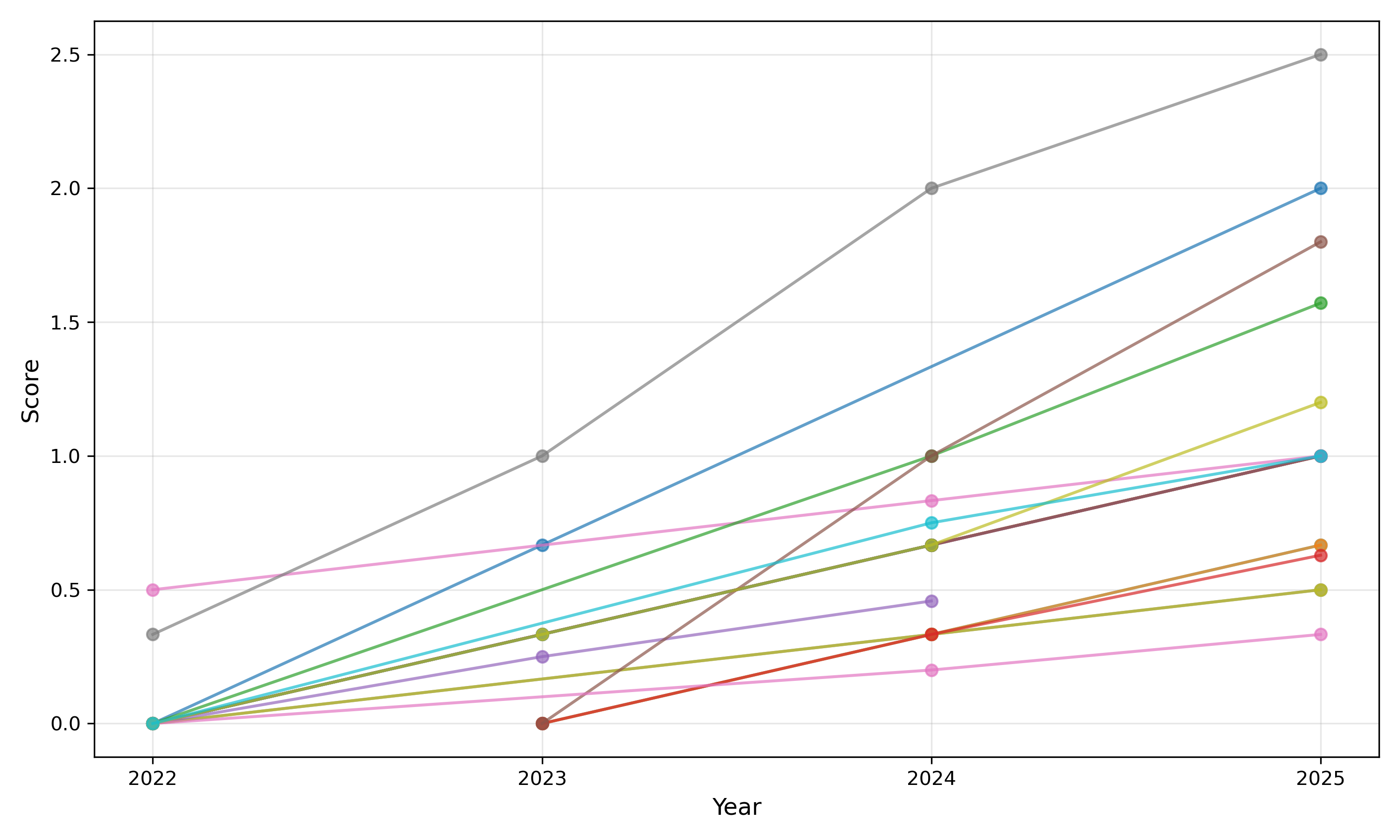}
\caption{Trajectories of the top 20 researchers in our rankings with the most statistically significant increasing score trend.}
\label{fig:emerging}
\end{figure*}

Figure~\ref{fig:emerging} shows the trajectories of the 20 researchers with the most statistically significant positive trends. It is worth noting that these individuals need not possess particularly high scores. One may begin with a modest score and remain below the top of the ranking while still exhibiting a strong and statistically significant upward trajectory, thereby qualifying as an emerging researcher under our definition. Such trajectories are consistent with researchers who are becoming increasingly successful at producing highly cited papers. Note that our definition of emerging researcher is not based on career stage or funding track record. Unlike many funding schemes, which define emerging researchers primarily through career age or eligibility criteria, the present approach identifies emergence through a statistically significant improvement in observed research performance. The term emerging is therefore used in a bibliometric rather than career stage sense.

\subsection{Institutional Strength Insights}
In this subsection, as another application we look at the insights our top 10\% papers in the \texttt{final\_top10percent\_papers\_by\_ASJC.xlsx} file give us for the strengths of affiliated institutions. The goal is not to rank institutions overall, but rather to identify the subject areas in which individual institutions make the strongest contributions to the highly cited tail of the literature. To do this, we assign the fractional scores of a paper to institutions through the authors and their normalised affiliations from the \texttt{final\_author\_scores.xlsx} file. In practice, this matching is done through the SCOPUS ID. We recall that a paper contributes a fractional score of $1/n$ to each ASJC, where $n$ is the number of ASJC codes assigned to the journal. We further divide this score by the number of Irish authors $m$ and assign the resulting fraction $1/(n \, m)$ to each author's normalised affiliated institution. Summing over each paper in the top 10\% in each ASJC gives us a score for the institution in each ASJC for each year in the window 2022--2025. To get an overall score for the institution, we average over the annual scores in the four year period. 

\begin{figure*}
\centering
\includegraphics[scale=0.6]{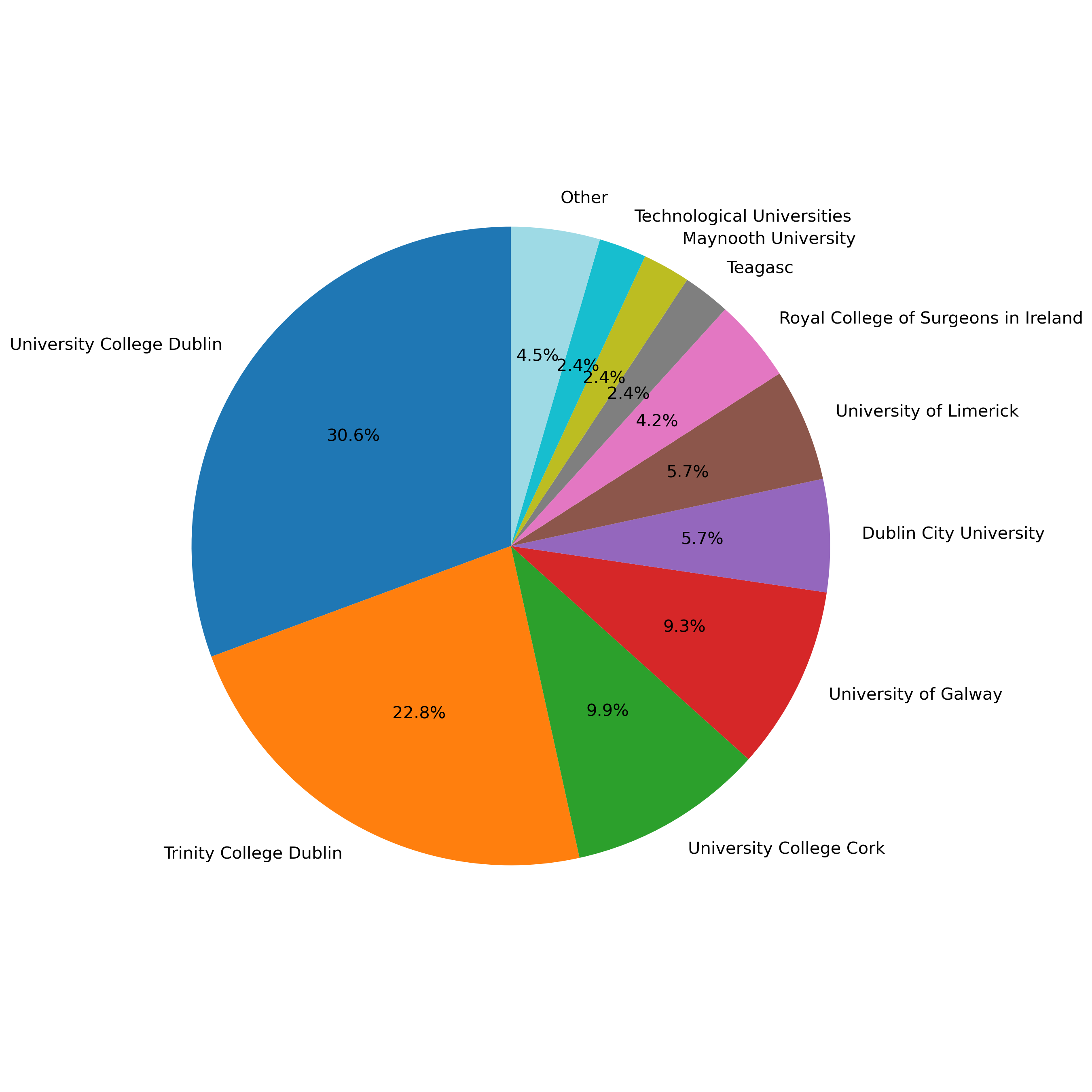}
\caption{Share of ASJC subjects where institutions achieve the highest score nationally. Technological Universities include Technological University Dublin, Atlantic Technological University and South East Technological University. Other includes Dublin Institute for Advanced Studies, hospitals and industry R\&D.}
\label{fig:asjc_wins}
\end{figure*}

The output of this exercise is a ranking of Irish institutions by ASJC, which given its length we present in the appendix. We note that the scores are dominated by the eight established universities (UCD, TCD, UCC, UG, DCU, UL, RCSI, MU). The remaining institutions comprise national laboratories (Teagasc, DIAS, Tyndall), technological universities, hospitals and industry. We retain the top 3 institutions in each subject, except in more specialised subjects where fewer institutions are active. In Fig. \ref{fig:asjc_wins}, restricting our attention to the top institution in each category we provide a breakdown of the number of subject areas in which the institution achieves the highest score nationally. We note that in specialised fields, the number of publications in the top 10\% is small, so it is typically in these fields where industry affiliations appear. 

Fig.~\ref{fig:asjc_wins} demonstrates that leadership in highly cited research remains concentrated within a relatively small number of institutions, in particular the established universities. Nevertheless, specialised institutions, hospitals, technological universities and industry also emerge as national leaders in selected subject areas. This highlights the diversity of the Irish research ecosystem and suggests that influential research lines are distributed across institutions.

\section{Overlap with Qualitative Peer Review}
In this section, we come to our second main result. Within university systems, securing competitive research funding is widely regarded as evidence of research excellence. At a very basic level, the awardee has written a proposal that has resonated with a panel of reviewers, be they generalists or experts. 
At the same time, funding bodies often pursue a range of objectives beyond scientific merit, including policy, societal, economic, or strategic goals. Nevertheless, the implicit assumption is that researchers selected through such processes will contribute to enhancing the reputation of their host institutions. Since funding is transferable, and institutions compete for researchers and grants, it is common that the awardee leverages a promotion. 

However, funding alone neither guarantees research output nor reputation enhancement. To attribute institutional reputation enhancement to the researchers, their research needs to attract attention from the wider academic community. Here we will demonstrate that funding awarded through qualitative peer review need not overlap with influential research in the tails of citation distributions. Bluntly put, funding success and being influential in one's field need not align. 

With our ranking of Irish researchers in place, we can turn our attention to the outcomes of qualitative peer review. We focus on Research Ireland's Frontiers for the Future Programme and the ERC's Starting, Consolidator and Advanced grants in the award years 2020--2025. Given our ranking spans 2022--2025, for earlier awardees our ranking provides an \textit{ex post} evaluation of performance, while for later awardees, our rankings provide \textit{ex ante} insights into the track record of the awardee leading up to the award. Our analysis does not distinguish before and after as the focus is not the project but the researcher. We will be interested in the score of the typical or median awardee. 

It should be noted that funding bodies support projects, not researchers, but the researcher is also evaluated during peer review and is expected to possess the capacity to execute the project successfully. We note that funding bodies award funding to researchers because they expect those researchers to exercise scientific judgement and leadership. Consequently, one would generally expect the most influential outputs of a successful project to bear clear evidence of the awardee's intellectual involvement. Thus, if a funded awardee does not produce any top 10\% papers, this constitutes evidence that neither the awardee nor the associated research lines have exerted substantial influence on the literature within our observation window extending to 2025. Admittedly, the connection is indirect, but one would nevertheless expect the performance of the awardee and the performance of the project to be positively correlated.

\subsection{Research Ireland}
Our approach here mirrors earlier analysis. All that is required is a SCOPUS ID for an awardee, which we assume to be unique, and this allows us to track the number of top 10\% papers the awardee has written in the period 2022--2025. It is important to bear in mind that we overcount papers when we extract the top 10\% in an ASJC, so our methods identify a pool of researchers connected to approximately 25\% of the papers. This is not particularly selective and this makes the results more surprising.  

Our approach for matching a SCOPUS ID to an awardee name is as follows. Research Ireland, formerly Science Foundation Ireland (SFI), provide the names of Frontiers for the Future Programme (FFP) awardees \footnote{\textit{www.sfi.ie/about-us/governance/open-data;www.researchireland.ie/about-us/governance-policies/open-data}
}, so we search for the awardee first in our ranking (\texttt{final\_author\_scores.xlsx}), next in our SCOPUS export (\texttt{ireland\_4years\_data.xlsx}), and if nothing appears, we resort to directly searching in SCOPUS by name and affiliation. The focus on FFP is justified as there is no targeted theme and the applicants submit proposals based on their own research expertise. Note, if an awardee is never active, namely never appears in our SCOPUS export with an Irish affiliation, they simply drop out of the analysis. It should be stressed here that SCOPUS assigns an ID to all authors on papers. In most cases, this coincides with the name of a researcher already in the database, but there are incidences where the variants of the same name appear with different SCOPUS IDs. SCOPUS allows for these profiles to be merged. We stress that while it is possible that some authors may have multiple IDs, their papers would still need to pass our top 10\% threshold to affect our results. As we shall see, the median Frontiers for the Future awardee lies well below the threshold for appearance in the ranking, so such effects are unlikely to alter the main conclusions.

Once we have the SCOPUS IDs for the awardees in each award year, we run them through each year of the ranking. As before, if the ID appears, we assign the score, if the ID does not appear but the awardee is active in Ireland we assign a score of zero, otherwise we assign NaN, where NaN drops out of all percentiles. In Table \ref{tab:ffp} we present a breakdown of the scores of the awardees for each year. The years represent the award years and not the evaluation years documented in FFP project codes. Irrespective of the year analysed, the median is zero, so the typical FFP awardee does not write a top 10\% paper in any year analysed. From scoring and active researchers, one can work out the percentage of awardees authoring a top 10\% paper, which when broken down by the cohorts in Table \ref{tab:ffp} and averaged over the years are respectively 37\%, 40\%, 26\%, 39\% and 30\%. The main takeaway is that irrespective of the year analysed, the typical awardee does not author a top 10\% paper that year even if publishing with an Irish affiliation. 

To place this result in context, in the previous section we showed that it was easy to construct researcher cohorts where the median researcher produces a greater return in terms of bibliometric performance. Note, if one lowers the threshold below top 10\% in an ASJC, corresponding to 25\% most influential papers, one may find that the typical Research Ireland awardee performs better. In \cite{2026:FWCI} it is shown that the majority of SFI projects have attracted greater than global median citations, but given that Ireland is expected to be above average, and our indicator is selective in the Irish context, the threshold here is higher.

\begin{table}[t] \centering \begin{tabular}{cccccc} 
\hline \hline
Year & Active & Scoring & Median & P75 & P90 \\
\hline
\multicolumn{6}{l}{\textbf{2020 Awardees (72)}}\\ 
2022 & 69 & 30 & 0 & 1.00 & 1.10 \\
2023 & 66 & 26 & 0 & 0.63 & 1.18 \\
2024 & 63 & 19 & 0 & 0.33 & 1.00 \\
2025 & 66 & 23 & 0 & 0.50 & 1.00 \\
\hline 
\multicolumn{6}{l}{\textbf{2021 Awardees (72)}}\\
2022 & 66 & 24 & 0 & 0.69 & 1.33 \\
2023 & 66 & 25 & 0 & 0.96 & 1.27 \\
2024 & 63 & 26 & 0 & 1.00 & 1.67 \\
2025 & 61 & 27 & 0 & 0.83 & 1.75 \\
\hline
\multicolumn{6}{l}{\textbf{2022 Awardees (81)}}\\
2022 & 71 & 19 & 0 & 0.25 & 1.00 \\
2023 & 65 & 12 & 0 & 0.00 & 0.50 \\
2024 & 70 & 18 & 0 & 0.19 & 1.18 \\
2025 & 69 & 23 & 0 & 0.50 & 2.00 \\
\hline 
\multicolumn{6}{l}{\textbf{2023 Awardees (68)}}\\ 
2022 & 62 & 24 & 0 & 1.00 & 2.40 \\
2023 & 57 & 20 & 0 & 0.67 & 1.70 \\
2024 & 58 & 24 & 0 & 0.84 & 2.00 \\
2025 & 62 & 26 & 0 & 1.00 & 1.50 \\
\hline
\multicolumn{6}{l}{\textbf{2024 (6) \& 2025 Awardees (53)}}\\
2022 & 47 & 16 & 0 & 0.67 & 1.52 \\
2023 & 50 & 17 & 0 & 0.50 & 2.00 \\
2024 & 54 & 13 & 0 & 0.00 & 1.00 \\
2025 & 52 & 14 & 0 & 0.28 & 1.00 \\
\hline
\multicolumn{6}{l}{\textbf{2020-2025 Awardees (352)}} \\
2022 & 315 & 113 & 0 & 1.00 & 1.50 \\
2023 & 304 & 100 & 0 & 0.50 & 1.27 \\
2024 & 308 & 100 & 0 & 0.50 & 1.25 \\
2025 & 310 & 113 & 0 & 0.75 & 1.01 \\
\hline
\hline
\end{tabular}
 \caption{From a cohort of 352 FFP awardees in the 2020--2025 window, we document active awardees, awardees appearing on the ranking with a non-zero score, and the ranking scores for the upper quartile (P75) and decile (P90). Given the median (P50) is zero, we omit lower percentiles.} 
 \label{tab:ffp}
 \end{table}

 From Table \ref{tab:ffp} we observe that the 2022 cohort, comprising 81 awardees, exhibits the lowest proportion of scoring researchers among all FFP cohorts at 26\%. This cohort is followed closely by the combined 2024+2025 cohort at 30\%. The same pattern is reflected in the upper quartile (P75) values. One possible explanation is that the 2022 call included ringfenced funding for researchers based in technological universities, meaning that applicants from established and technological universities did not compete directly. Seventeen technological university awards were made in the 2021 call. Across all Frontiers cohorts, a total of 24 awards were made to technological university researchers, with the remaining seven awards distributed across the other years. Given the visibly weaker performance of the 2022 cohort, it is therefore natural to ask whether the inclusion of technological university awardees might explain the observed reduction in bibliometric performance.

 To investigate this possibility, we split the 352 Frontiers awardees into 24 technological university awardees and 328 awardees from established universities. The resulting statistics are shown in Table \ref{tab:tu_versus_u}. While the technological university cohort generally exhibits lower P75 values, the sample is extremely small, with only 19--21 active researchers in any given year, making upper-tail statistics particularly unstable. More importantly, the dominant pattern of the FFP remains unchanged after decomposition. Both cohorts exhibit median scores of zero throughout the observation period, and the overall differences between the two groups are modest. Consequently, there is little evidence that the weak bibliometric performance of the FFP awardees in 2022 can be attributed to the inclusion of technological university researchers.  Overall, we find little difference between established university and technological university awardees selected by SFI/Research Ireland. 

\begin{table}[t] \centering \begin{tabular}{cccccc} 
\hline \hline
Year & Active & Scoring & Median (P50) & P75 & P90 \\
\hline
\multicolumn{6}{l}{\textbf{Technological University Awardees (24)}}\\ 
2022 & 21 & 7 & 0 & 1.00 & 1.14 \\
2023 & 20 & 4 & 0 & 0.00 & 0.55 \\
2024 & 19 & 6 & 0 & 0.33 & 1.80 \\
2025 & 20 & 9 & 0 & 0.75 & 2.00 \\
\hline 
\multicolumn{6}{l}{\textbf{University Awardees (328)}}\\
2022 & 293 & 105 & 0 & 1.00 & 1.50 \\
2023 & 282 & 95 & 0 & 0.50 & 1.33 \\
2024 & 287 & 93 & 0 & 0.50 & 1.25 \\
2025 & 288 & 103 & 0 & 0.75 & 1.00 \\
\hline
\hline
\end{tabular}
\caption{Same as Table \ref{tab:ffp} but the 352 Frontiers awardees split between 24 technological university awardees and 328 established university awardees. }
\label{tab:tu_versus_u}
\end{table}

\subsection{European Research Council}

Having examined Research Ireland's FFP, it is instructive to consider the overlap between ERC awardees and our ranking. Both schemes rely primarily on qualitative peer review rather than quantitative indicators, and both draw largely from the same Irish research system. Important differences nevertheless exist. The ERC is focused exclusively on scientific excellence, evaluations are conducted by international subject experts, and the awards provide substantially greater resources for building research teams. Consequently, one might reasonably expect ERC awardees to outperform FFP awardees on our top 10\% indicator. 

\begin{table}[t] \centering 
\begin{tabular}{cccccc}
\hline\hline
Year & Active & Scoring & Median & P75 & P90 \\
\hline
\multicolumn{6}{l}{\textbf{Starting Awardees (45)}}\\
2022 & 23 & 12 & 0.17 & 0.88 & 1.80 \\
2023 & 27 & 11 & 0.00 & 0.88 & 1.63 \\
2024 & 27 & 10 & 0.00 & 0.75 & 1.85 \\
2025 & 32 & 12 & 0.00 & 1.00 & 1.74 \\
\hline
\multicolumn{6}{l}{\textbf{STEM Starting Awardees (25)}}\\
2022 & 14 & 7 & 0.12 & 1.00 & 2.21 \\
2023 & 16 & 8 & 0.17 & 1.37 & 1.67 \\
2024 & 18 & 6 & 0.00 & 0.88 & 1.83 \\
2025 & 22 & 9 & 0.00 & 1.00 & 1.98 \\
\hline
\multicolumn{6}{l}{\textbf{Consolidator Awardees (44)}}\\
2022 & 28 & 10 & 0.00 & 0.83 & 1.00 \\
2023 & 30 & 12 & 0.00 & 0.63 & 1.00 \\
2024 & 26 & 10 & 0.00 & 0.79 & 1.50 \\
2025 & 33 & 10 & 0.00 & 0.50 & 1.50 \\
\hline
\multicolumn{6}{l}{\textbf{STEM Consolidator Awardees (17)}}\\
2022 & 14 & 6 & 0.00 & 0.77 & 0.95 \\
2023 & 16 & 7 & 0.00 & 0.54 & 1.10 \\
2024 & 17 & 8 & 0.00 & 1.00 & 1.40 \\
2025 & 16 & 8 & 0.10 & 1.50 & 2.04 \\
\hline
\multicolumn{6}{l}{\textbf{Advanced Awardees (21)}}\\
2022 & 13 & 7 & 0.25 & 1.00 & 1.47 \\
2023 & 14 & 7 & 0.33 & 1.00 & 2.40 \\
2024 & 9 & 5 & 0.75 & 1.20 & 3.12 \\
2025 & 15 & 4 & 0.00 & 0.08 & 2.74 \\
\hline
\multicolumn{6}{l}{\textbf{STEM Advanced Awardees (8)}}\\
2022 & 7 & 4 & 0.25 & 1.29 & 3.34 \\
2023 & 8 & 6 & 0.92 & 1.50 & 3.20 \\
2024 & 6 & 5 & 0.98 & 2.36 & 3.67 \\
2025 & 7 & 3 & 0.00 & 0.58 & 2.20 \\
\hline
\hline
\end{tabular}
\caption{From a cohort of 110 ERC awardees in the 2020--2025 window, we document active awardees, awardees appearing on the ranking with a non-zero score, and the ranking scores for the upper quartile (P75) and decile (P90). Given the median (P50) is zero, we omit lower percentiles. For starting, consolidator and advanced we remove AHSS awardees to focus on STEM and get a better comparison with Table \ref{tab:ffp}.} \label{tab:erc}
\end{table}

Under the same assumptions, in Table \ref{tab:erc} we provide the analogous numbers for ERC Starting, Consolidator and Advanced grants. Given that FFP is essentially a STEM funding call, we further isolate STEM ERC awardees within each cohort. Some of the motivation is to facilitate like-for-like comparison, but motivation also comes from the understanding that citations may be a better indicator of influence in STEM than Arts, Humanities, Social Science (AHSS) subjects. To further aid comparison, we have restricted our attention to the same award years 2020--2025, but given the small number of ERC awards, we have combined all years in one cohort. For this reason, the numbers in each cohort are best compared to the performance of 325 FFP awardees over the 2020--2025 time window. To ease comparison, we note that $113/315 \approx 36\%$, $100/304 \approx 33 \%$, $100/308 \approx 32 \%$ and $113/310 \approx 36\%$ of FFP awardees appear on the ranking in 2022, 2023, 2024 and 2025, respectively. 

One interesting observation is that the Starting grant cohort appears to marginally outperform the Consolidator cohort both with and without AHSS awardees. Regardless, the performance of the typical Starting or Consolidator awardee is either zero or close to zero in all years. In line with expectations, we see that removing AHSS awardees increases the percentage of scoring relative to active awardees. Only when one gets to Advanced grants does the median become non-zero, where once again STEM awardees drive the signal. The biggest difference between ERC Starting/Consolidator grant awardees and FFP awardees is restricting to STEM, the percentage of active researchers appearing on our ranking approaches close to $50 \%$. More concretely, the numbers for STEM Starting awardees are $7/14 = 50 \%$, $8/16 = 50 \%$, $6/18 \approx 33 \%$ and $41 \%$, while for STEM Consolidator awardees we have $6/14 \approx 43\%$, $7/16 \approx 44\%$, $8/17 \approx 47\%$ and $8/16 = 50 \%$. These numbers demonstrate that a greater percentage of active Starting and Consolidator ERC awardees make our ranking compared to FFP awardees. The performance difference between FFP and ERC Starting and Consolidator awardees is of the order of 10\%. Given the differences in programme objectives, panel expertise and resources committed, this is a surprisingly modest separation. 

Nevertheless, given the performance of the typical awardee, we conclude that ERC peer review, like FFP peer review, identifies a substantially different cohort of researchers from that identified through persistent participation in the highly cited tail of the literature.

\subsection{Discussion}

The main takeaway from our analysis is that a top 10\% bibliometric indicator, one retaining 25\% of papers, and qualitative peer review identify substantially different cohorts of researchers. The strongest overlap occurs for ERC Advanced grant awardees, but even there the typical awardee stands out less than the typical researcher appearing in our ranking each year. It is important to note that funding agencies often showcase exceptional projects and individual success stories when demonstrating programme impact. While such cases are undoubtedly valuable, they provide limited insight into the performance of the typical awardee selected through the funding process. For this reason, median outcomes are generally more informative when evaluating the effectiveness of a funding programme as a whole. 

This motivates examining programme objectives through the lens of median rather than exceptional outcomes. Note, prizes, awards, patents and policy citations do not occur in sufficient volumes to characterise median outcomes; only papers occur in sufficient numbers to allow one to evaluate the outputs of the median researcher. Viewed through this lens, both Research Ireland and ERC peer review identify cohorts whose typical member contributes only modestly to the highly cited tail of the literature. By contrast, the bibliometrically defined cohorts identified in the previous section exhibit persistent participation in the highly cited literature even when selected using only a single year of data.

Historically, Research Ireland's FFP funding call started off as an SFI call before SFI was rebranded as Research Ireland in an amalgamation with the Irish Research Council. Nevertheless, the FFP funding call spans both organisations and its evaluation procedures and cultural makeup are expected to remain the same. While Research Ireland does not explicitly rank or weight its objectives \footnote{www.researchireland.ie/funding/frontiers}, programme outcomes suggest that some objectives are more successfully realised than others. For example, the objective concerning support for excellent women researchers, ranked third in the list of objectives, appears to have translated into close to gender parity among awardees, whereas the objective concerning development of research capacity within the technological university sector, which appears seventh, has translated into a substantially smaller share of awards even with funding ringfencing, approximately 7\%. This suggests that the practical emphasis of the programme is unlikely to be uniform across all objectives. 

From the ten FFP objectives, our results are most relevant to objectives 1), 4) and 6): 

\begin{quote}

\textit{1) To build research capacity, expertise and reputation through funding excellent scientific research}

\textit{4) To retain and improve Ireland’s reputation as a place to carry out excellent research}

\textit{6) To allow Ireland-based researchers to win top-tier international prizes}

\end{quote} 

Objective 1) is open to interpretation. Research capacity and expertise may be developed without enhancing international research reputation. On the other hand, international reputation is enhanced through various channels, including highly cited papers, prestigious grants, scientific prizes, patents/commercialisation, policy citations, etc. Nevertheless, once one moves away from publications, the number of observable outputs decreases rapidly. The rareness of these other outcomes relative to highly cited papers makes the latter the most direct and widely applicable route to enhance international academic reputation. That being said, funding agencies operate within a domestic policy environment, where reputation may be understood in terms of national visibility, stakeholder engagement, or institutional development rather than international influence on the research literature. Under such an interpretation, there is less conflict between objective 1) and our findings that the typical FFP awardee writes no top 10\% papers each year. 

Objective 4) seeks to improve Ireland's reputation as a place to carry out excellent research. Under the interpretation adopted here, influence on the highly cited tail of the literature provides a direct indicator of international research reputation. Nevertheless, the median FFP awardee exhibits a score of zero throughout the observation period, providing little evidence  that the typical awardee enhances Ireland's international research reputation through influence on the scientific literature. Note, this is a relative statement, the typical awardee stands out much less than the typical researcher on our ranking. Objective 6) seeks to enable researchers to win top-tier international prizes. Such prizes are generally awarded to exceptional researchers whose influence on a field has been sustained over many years. While prizes themselves are inherently rare and never median outcomes, they nevertheless emerge from a broader population of influential researchers. The weak performance of the median FFP awardee, which is more reflective of the majority, raises questions about whether the pool of researchers is sufficiently deep to support this objective. 

More recently, Research Ireland has started an Investigators programme \footnote{www.researchireland.ie/funding/investigators}, which may be regarded as a successor to FFP. The FFP objectives have been streamlined with less emphasis on policy and more focus on research excellence. The most relevant objectives in Investigators are the following: 

\begin{quote}
\textit{1) To enable excellent researchers to undertake cutting-edge research to push the boundaries of knowledge}
    
\textit{3) To increase Ireland’s capacity and reputation, and those of its research-performing institutions, for excellent research across all disciplines }
\end{quote}

What is striking here is that objective 1) closely mirrors the mission of the ERC, namely to pursue scientific excellence through frontier research without an explicit commitment to generate impact. This is essentially a commitment to support researchers tackling important open problems as judged through expert peer review. As a consequence, the role of subject specific expertise becomes increasingly important, since evaluating frontier research necessarily requires reviewers capable of assessing the significance of problems. This brings Research Ireland more in line with the research excellence focus of CoARA \cite{coara:2022}. Nevertheless, enhancing Ireland's reputation is once again a focus, where it is elevated from objective 4) in FFP to objective 3) in Investigators.  Given our ranking, once the first cohort of Investigator awardees is announced we will be in a position to examine the top 10\% paper track record of the typical awardee. Assuming Research Ireland truly embraces expert peer review, coupled with the focus on research excellence, it is reasonable to assume the 10\% gap with the typical ERC Starting/Consolidator awardee gets narrowed. Whether this is reflected in the median awardee remains an empirical question that can be tested. We do not have to wait for outcomes and with the imminent release of the first batch of Investigator awardees, one can comment on their track record leading up to the award.  

Finally, this brings us to the ERC, where the central objective is \footnote{erc.europa.eu/about-erc/erc-glance}: 

\begin{quote}
\textit{to encourage the highest quality research in Europe through competitive funding and to support investigator-driven frontier research across all fields, based on scientific excellence}
\end{quote}

It cannot be stressed enough here that neither this objective nor objective 1) from the Investigator programme commit to generating impact. At this point an analogy may be useful. One may think of frontier research as opening gates into new pastures in the hope that the wider research community subsequently traverses them. Therein lies the impact. Within this analogy, bibliometric indicators such as our top 10\% metric track researchers whose work successfully attracts the attention of the wider community and thereby influences which directions are subsequently explored. In contrast, the ERC concerns itself with gate opening. Irrespective of the level of the award, be it Starting, Consolidator or Advanced, at least within the time windows analysed and the confines of Irish awardees, our findings are consistent with the herd not traversing a gate, even if an Irish awardee has been successful in opening one. 

It is plausible that the herd traverses the gate later, but the rareness of sleeping beauty papers \cite{vanRaan:2004, Ho:2017, Teixeira:2017} suggests that this is unlikely to happen in the majority of cases. Even if the herd traverses the gate much later, will the herd make the connection to earlier work? We note that Horndeski wrote down a modified gravity theory in 1974 \cite{Horndeski:1974wa} but subsequently left the field. The theory was independently rediscovered in 2011 \cite{Kobayashi:2011nu} at a time when the research community was interested in modified gravity theories \cite{Clifton:2011jh}. Although Horndeski eventually received credit for his work almost four decades later, providing a quintessential example of a sleeping beauty paper, the subsequent development of the field was driven by renewed interest in modified gravity rather than engagement with the original paper. For this reason, it is not enough to have the right idea, but also to have the right idea at a time when the community is interested in the problem. Gate opening does not lead to impact unless the herd traverses the gate. Note, while ERC funding is seen as competitive and prestigious, if our Irish findings are representative of the ERC programme as a whole, a rethink on the typical impact may be warranted.  

As a further aside, in funding circles, it is common to cherry pick exceptional awardees or outliers in the data to justify funding calls. However, given a very large budget, i.e. the ERC, any sufficiently large funding programme will produce outstanding outliers. From a statistical perspective, it is not so meaningful to evaluate a programme on the upper percentiles and it is far more instructive to focus on the performance of the typical or median awardee as this better reflects the performance of the majority of funded researchers. Note, without a metric or measure one cannot define a median. Our paper takes an important step towards providing greater transparency on the typical awardee's performance. In terms of a policy recommendation, our results suggest that a bibliometric pre-screening stage based on prior research performance should identify a cohort with higher subsequent influence on the field than peer review alone. 

We are conscious that mining past trajectories through bibliometrics and assessing research through qualitative peer review are philosophically different. The biggest difference is that the former is regarded as retrospective whereas the latter attempts to make predictions. Nevertheless, as we have shown, bibliometric indicators can make predictions based on historical data and these predictions ensure that cohorts identified this way already outperform qualitative peer review awardees when it comes to influence in the literature. This suggests that one could pre-screen for researchers writing papers in the tails of field weighted citations, allow them to submit proposals, review everything qualitatively through peer review and improve overall influence in the field with respect to the median. This constitutes a plausible policy recommendation emerging from the present analysis. 

While data-driven pre-screening is now central to professional sports, where the goal is to identify players in the data before going to the expense of watching them play, one could attempt to argue that research is not competitive. Nevertheless, at the national level research is competitive, where national research reputation, the ability to attract talent and foreign direct investment are all connected to international visibility and influence. Based on our results, data-driven pre-screening appears to add predictive information beyond qualitative peer review. This shifts the burden of proof to those arguing that such information should be ignored. We stress that data has not replaced traditional scouts in professional sports, simply given them access to more information to make better informed decisions. 

\section{Conclusions}

Working nationally in a funding jurisdiction, Ireland, we introduced a top 10\% metric for field weighted citations through journal ASJC codes and propagated the scores to all Irish co-authors of the paper. We restricted first or corresponding authorship to Ireland to ensure more of the research was conducted locally. Since each paper may be assigned to multiple ASJC codes through its journal, our metric is less selective and retains approximately 25\% of papers in any year. Although the top 10\% ASJC threshold is an arbitrary choice, our motivation was to identify papers that are in principle on track to be in the tails of citation distributions with associated greater demonstrable influence on the field. Our methods made no distinction within the top 10\%, but one could easily introduce a gradient to allow greater distinction. The final author score is easily interpreted as the number of ``whole" papers a researcher co-authors in the top 10\% of field weighted citations across all subfields, where purely collaborative roles are not penalised. 

It should also be borne in mind that academics are employed in part to undertake research. Since a substantial proportion of this activity is publicly funded, there is a legitimate public interest in understanding the extent to which publicly funded research attracts attention from the wider scholarly community. For reasons of transparency, reproducibility and public scrutiny, we therefore release our rankings into the public domain alongside this paper. Beyond these considerations, influential papers constitute one of the most visible and numerous mechanisms through which universities build international research reputation. Consequently, researchers who repeatedly contribute to the highly cited tail of the literature are of particular interest to universities, funding agencies and the wider public.

Based on four years of researcher rankings 2022--2025, we demonstrated that authoring papers in the tails of citation distributions is a persistent researcher characteristic provided researchers author strictly more than one top 10\% paper annually. Even though we apply our indicator to \textit{ex post} research evaluation where one may want a longer window \footnote{Note, there is no reason to expect the performance of the typical researcher changes much over a longer time window.}, for \textit{ex ante} research assessment, a short time window leading up to funding application is preferable. Even in Ireland, this identifies a cohort of no less than 500 researchers a year, and for up to three subsequent years - the extent of our data - the typical or median researcher in the cohort succeeds to co-author a paper that is in the top 10\% of field weighted citations in a subfield, leading to a non-zero median score. Against a backdrop where up to 4,000 domestic researchers co-author a top 10\% paper each year, we find that the median Research Ireland FFP or ERC Starting and Consolidator grant awardee selected through qualitative peer review does not appear. 

We conclude that bibliometrics tracking influence in the field and qualitative peer review select substantially different cohorts of researchers. Throughout, it is worth emphasising that our ERC awardee samples are small, and they are made even smaller once AHSS awardees are removed. For this reason, it would be interesting to scale the study to a large sample of ERC awardees. For small nations such as Ireland, the observation that the typical awardee authors no top 10\% papers sits uneasily with objectives centred on enhancing the international reputation of either the nation or the host universities. Note, internationally researchers engage with titles, abstracts, authors, institutions, but one rarely reads acknowledgements to see how research is funded, so the prestige of winning an award, even an ERC award, is not expected to travel unless the research community engages with research outputs. Although we only study a time window equivalent to a traditional funding cycle, it is plausible that papers attract greater interest later, but the rareness of sleeping beauty papers is well documented \cite{vanRaan:2004, Ho:2017, Teixeira:2017}.   

While this paper focuses on research influence as measured through citations, the underlying philosophy is broader. Any publicly observable indicator of research impact can, in principle, be incorporated into a similar framework. We plan on extending the analysis to mine patents, collaborations with industry, policy citations through \textit{Overton} and potentially even \textit{altmetrics} to identify researchers with a track record of economic, policy and societal impacts in publicly available databases. This will allow us to comment further on the performance of typical awardees. One issue one will encounter is that such outputs may be rarer and the time lag between research and output may be longer. 

We note that a focus on objective outputs is arguably more scientific than \textit{ex ante} self-reported impact statements, e. g. narrative CVs \footnote{Recent restrictions on hyperlinks and URLs \textit{www.researchireland.ie/funding-policies/narrative-cv}, with a view to improving fairness, create a tension with normal academic practice, whereby academics are expected to provide supporting evidence. This ultimately decreases the information accessible to reviewers and increases noise.}, and \textit{ex post} cases studies allowing success to be defined retrospectively where comparison between funded awards is near impossible. Objective indicators are observable, independently verifiable, reproducible and comparable, whereas narrative approaches rely more heavily on self-reporting and subjective interpretation. It is puzzling that research evaluation can remain almost exclusively qualitative in a world that has moved to data informed decisions, where it is worth emphasising that CoARA allows for supporting metrics \cite{coara:2022}.  

\section*{Acknowledgements}
We acknowledge the use of Microsoft's Copilot in informing the literature review, writing Python code that was validated and executed locally, and improving presentation or wording in the manuscript. Hirushan Sajindra is supported by RISE@ATU. RISE@ATU is co-funded by the Government of Ireland and the European Union through the ERDF Northern \& Western Regional Programme 2021-2027.

\bibliography{refs}

\appendix 

\section{Appendix: Institutional Strengths}
In this appendix we document the research strengths of the Irish institutions based on top 10\% papers in ASJC subfields. We restrict our attention to the three institutions in each ASJC with the greatest score and report fewer institutions in specialised subfields with lower activity levels. In the following table we abbreviate the affiliations as follows: ATU (Atlantic Technological University), DCU (Dublin City University), DIAS (Dublin Institute for Advanced Studies), DkIT (Dundalk Institute of Technology), MTU (Munster Technological University), MU (Maynooth University), RCSI (Royal College of Surgeons in Ireland), SETU (South East Technological University), TCD (Trinity College Dublin), TUD (Technological University Dublin), TUS (Technological University of the Shannon), UCC (University College Cork), UCD (University College Dublin), UG (University of Galway) and UL (University of Limerick)

\begin{table*} 
\centering 
\begin{tabular*}{\textwidth}{@{\extracolsep{\fill}} lll}
\hline
\hline
ASJC & Subject Area & Top Institutions \\
\hline 
\multicolumn{3}{l}{\textbf{Multidisciplinary}}\\
1000 & Multidisciplinary & TCD, UCC, UCD \\
\multicolumn{3}{l}{\textbf{Agricultural and Biological Sciences}}\\
1100 & General Agricultural and Biological Sciences & UCD, UCC, ATU \\
1101 & Agricultural and Biological Sciences (miscellaneous) & Teagasc, UCD, UG \\
1102 & Agronomy and Crop Science & UCD, Teagasc, MTU \\
1103 & Animal Science and Zoology & Teagasc, UCD, UCC \\
1104 & Aquatic Science & UCD, UG, TCD \\
1105 & Ecology, Evolution, Behavior and Systematics & TCD, UCD, UCC \\
1106 & Food Science & Teagasc, UCD, UCC \\
1107 & Forestry & UCD, UCC, TCD \\
1108 & Horticulture & UCD, TCD, Teagasc \\
1109 & Insect Science & Teagasc, ClonBio Group LTD, MU \\
1110 & Plant Science & UCD, UCC, Teagasc \\
1111 & Soil Science & UCD, Teagasc, UG \\
\multicolumn{3}{l}{\textbf{Arts \& Humanities}}\\
1200 & General Arts and Humanities & UCD, UCC, UG \\
1201 & Arts and Humanities (miscellaneous) & DCU, UL, UCC \\
1202 & History & UCD, UCC, DCU \\
1203 & Language and Linguistics & DCU, UCD, TCD \\
1204 & Archeology & UCD, UCC, MU \\
1205 & Classics & TCD, UCD, UG \\
1206 & Conservation & MU, TCD, UCD \\
1207 & History and Philosophy of Science & TCD, UG, UCD \\
1208 & Literature and Literary Theory & DCU, UCD, TCD \\
1209 & Museology & MU, TCD, DCU \\
1210 & Music & UL, Mary Immaculate College, DCU \\
1211 & Philosophy & UCD, TCD, UG \\
1212 & Religious studies & DCU, UCD, Mary Immaculate College \\
1213 & Visual Arts and Performing Arts & TCD, MU, DCU \\
\multicolumn{3}{l}{\textbf{Biochemistry, Genetics and Molecular Biology}}\\
1300 & General Biochemistry, Genetics and Molecular Biology & TCD, UCD, UCC \\
1301 & Biochemistry, Genetics and Molecular Biology (miscellaneous) & UG, UCC, Teagasc \\
1302 & Aging & TCD, UCD, UCC \\
1303 & Biochemistry & TCD, UL, UCD \\
1304 & Biophysics & TCD, UCD, UCC \\
1305 & Biotechnology & UCD, UG, UL \\
1306 & Cancer Research & UCD, St Vincent's Hospital, TCD \\
1307 & Cell Biology & TCD, UCC, UCD \\
1308 & Clinical Biochemistry & UCD, UCC, UL \\
1309 & Developmental Biology & TCD, UCD, UCC \\
1310 & Endocrinology & UCD, RCSI, UCC \\
1311 & Genetics & UCD, Teagasc, UCC \\
1312 & Molecular Biology & TCD, UCD, RCSI \\
1313 & Molecular Medicine & UCC, UG, TCD \\
1314 & Physiology & UCD, DCU, TCD \\
1315 & Structural Biology & UL, UCC, Teagasc \\
\multicolumn{3}{l}{\textbf{Business, Management and Accounting}}\\
1400 & General Business, Management and Accounting & TCD, UG, DCU \\
1401 & Business, Management and Accounting (miscellaneous) & TCD, DCU, UCC \\
1402 & Accounting & UCD, DCU, TCD \\
1403 & Business and International Management & DCU, MU, UL \\
1404 & Management Information Systems & UG, DCU, UCD \\
1405 & Management of Technology and Innovation & MU, UCD, UG \\
1406 & Marketing & TCD, UCD, MU \\
1407 & Organizational Behavior and Human Resource Management & UL, UCC, TCD \\
1408 & Strategy and Management & UCD, DCU, UG \\
1409 & Tourism, Leisure and Hospitality Management & DCU, TUD, UG \\
1410 & Industrial relations & UL, UCD, Nitro Software \\
\multicolumn{3}{l}{\textbf{Chemical Engineering}}\\
1500 & General Chemical Engineering & UCC, UL, Teagasc \\
1501 & Chemical Engineering (miscellaneous) & TUD, UG, UCD \\
1502 & Bioengineering & UG, UCD, RCSI \\
1503 & Catalysis & TCD, UL, UCD \\
1504 & Chemical Health and Safety & UCD, UG, DCU \\
1505 & Colloid and Surface Chemistry & UL, UCD, DCU \\
\end{tabular*}
\end{table*}

\begin{table*} 
\centering
\begin{tabular*}{\textwidth}{@{\extracolsep{\fill}} lll}
\hline
\hline
ASJC & Subject Area & Top Institutions \\
\hline 
1506 & Filtration and Separation & UG, UCD \\
1507 & Fluid Flow and Transfer Processes & UCD, ATU, UG \\
1508 & Process Chemistry and Technology & UG, UL, UCD \\
\multicolumn{3}{l}{\textbf{Chemistry}}\\
1600 & General Chemistry & TCD, UL, UG \\
1601 & Chemistry (miscellaneous) & UCD, TUS, DCU \\
1602 & Analytical Chemistry & UCD, Teagasc, UL \\
1603 & Electrochemistry & TCD, UL, UCC \\
1604 & Inorganic Chemistry & UCD, UL, DCU \\
1605 & Organic Chemistry & UCD, UG, RCSI \\
1606 & Physical and Theoretical Chemistry & UCD, TCD, UL \\
1607 & Spectroscopy & UCD, TCD, TUD \\
\multicolumn{3}{l}{\textbf{Computer Science}}\\
1700 & General Computer Science & UCD, UG, UL \\
1701 & Computer Science (miscellaneous) & ATU, UCD, DCU \\
1702 & Artificial Intelligence & UCD, UG, TCD \\
1703 & Computational Theory and Mathematics & UCD, UG, TUD \\
1704 & Computer Graphics and Computer-Aided Design & TCD, UCD, UCC \\
1705 & Computer Networks and Communications & UCD, TCD, UG \\
1706 & Computer Science Applications & UCD, UG, DCU \\
1707 & Computer Vision and Pattern Recognition & DCU, TCD, UCD \\
1708 & Hardware and Architecture & UCD, UG, TCD \\
1709 & Human-Computer Interaction & TCD, UCD, DCU \\
1710 & Information Systems & UG, UCD, TCD \\
1711 & Signal Processing & UCD, DCU, TCD \\
1712 & Software & UCD, UG, TCD \\
\multicolumn{3}{l}{\textbf{Decision Sciences}}\\
1800 & General Decision Sciences & DCU, UCD, MU \\
1801 & Decision Sciences (miscellaneous) & UCC, UCD, UL \\
1802 & Information Systems and Management & UCD, UG, MU \\
1803 & Management Science and Operations Research & UCD, DCU, TCD \\
1804 & Statistics, Probability and Uncertainty & MU, UG, TCD \\
\multicolumn{3}{l}{\textbf{Earth and Planetary Sciences}}\\
1900 & General Earth and Planetary Sciences & UG, UCD, MU \\
1901 & Earth and Planetary Sciences (miscellaneous) & UG, UCD, TCD \\
1902 & Atmospheric Science & UG, MU, TCD \\
1903 & Computers in Earth Sciences & MU, Ubotica Technologies, UCD \\
1904 & Earth-Surface Processes & MU, UCC, UG \\
1905 & Economic Geology & UCD, UCC, TCD \\
1906 & Geochemistry and Petrology & DIAS, TCD, UCD \\
1907 & Geology & TCD, UCC, UCD \\
1908 & Geophysics & DIAS, UCD, TCD \\
1909 & Geotechnical Engineering and Engineering Geology & UCC, UCD, TCD \\
1910 & Oceanography & TCD, UCC, UCD \\
1911 & Paleontology & TCD, UCD, UCC \\
1912 & Space and Planetary Science & TCD, ATU, DIAS \\
1913 & Stratigraphy & Teagasc, UG, DCU \\
\multicolumn{3}{l}{\textbf{Economics, Econometrics and Finance}}\\
2000 & General Economics, Econometrics and Finance & UCD, TCD, DCU \\
2001 & Economics, Econometrics and Finance (miscellaneous) & UCD, TCD, DCU \\
2002 & Economics and Econometrics & TCD, UCD, DCU \\
2003 & Finance & TCD, DCU, UCD \\
\multicolumn{3}{l}{\textbf{Business, Management and Accounting}}\\
2100 & General Energy & TCD, UCC, DCU \\
2101 & Energy (miscellaneous) & UCC, UG, UL \\
2102 & Energy Engineering and Power Technology & UCC, TCD, UG \\
2103 & Fuel Technology & UG, UCC, UCD \\
2104 & Nuclear Energy and Engineering & UCC, UCD, MU \\
2105 & Renewable Energy, Sustainability and the Environment & TCD, UCC, UCD \\
\multicolumn{3}{l}{\textbf{Engineering}}\\
2200 & General Engineering & UCD, ATU, UG \\
2201 & Engineering (miscellaneous) & ATU, TCD, UCC \\
2202 & Aerospace Engineering & UCD, MU, UCC \\
2203 & Automotive Engineering & UL, UCD, UG \\
2204 & Biomedical Engineering & UCD, UG, RCSI \\
2205 & Civil and Structural Engineering & UCD, UG, TCD \\
2206 & Computational Mechanics & TCD, UCD, RCSI \\
\end{tabular*}
\end{table*}

\begin{table*} 
\centering
\begin{tabular*}{\textwidth}{@{\extracolsep{\fill}} lll}
\hline
\hline
ASJC & Subject Area & Top Institutions \\
\hline 
2207 & Control and Systems Engineering & UCD, TCD, UCC \\
2208 & Electrical and Electronic Engineering & UCD, UCC, DCU \\
2209 & Industrial and Manufacturing Engineering & UCD, UG, UL \\
2210 & Mechanical Engineering & UG, UCD, TCD \\
2211 & Mechanics of Materials & TCD, UG, UL \\
2212 & Ocean Engineering & UCD, MU, UG \\
2213 & Safety, Risk, Reliability and Quality & UCD, UG, TCD \\
2214 & Media Technology & UG, UL, DCU \\
2215 & Building and Construction & UCD, UCC, TUD \\
2216 & Architecture & UCD, Integrated Environmental Solutions R\&D, UCC \\
\multicolumn{3}{l}{\textbf{Environmental Science}}\\
2300 & General Environmental Science & TCD, UG, UCD \\
2301 & Environmental Science (miscellaneous) & UCD, TCD, UG \\
2302 & Ecological Modeling & UG, DkIT, MU \\
2303 & Ecology & UCD, TCD, UCC \\
2304 & Environmental Chemistry & UG, TCD, UCD \\
2305 & Environmental Engineering & UG, UCD, UCC \\
2306 & Global and Planetary Change & TCD, UG, UCD \\
2307 & Health, Toxicology and Mutagenesis & UCD, UG, TCD \\
2308 & Management, Monitoring, Policy and Law & UCD, UG, TCD \\
2309 & Nature and Landscape Conservation & TCD, UCC, UCD \\
2310 & Pollution & UG, UCD, UCC \\
2311 & Waste Management and Disposal & UCD, UG, UCC \\
2312 & Water Science and Technology & UG, ATU, TCD \\
\multicolumn{3}{l}{\textbf{Environmental Science}}\\
2400 & General Immunology and Microbiology & UCC, TCD, UCD \\
2401 & Immunology and Microbiology (miscellaneous) & Teagasc, ATU, UG \\
2402 & Applied Microbiology and Biotechnology & UCD, UCC, Teagasc \\
2403 & Immunology & TCD, UCC, RCSI \\
2404 & Microbiology & UCC, Teagasc, UCD \\
2405 & Parasitology & UG, UL, UCC \\
2406 & Virology & UCD, UG, TUD \\
\multicolumn{3}{l}{\textbf{Environmental Science}}\\
2500 & General Materials Science & TCD, UG, UCD \\
2501 & Materials Science (miscellaneous) & UCD, SETU, MTU \\
2502 & Biomaterials & TCD, UCD, UL \\
2503 & Ceramics and Composites & UL, UG, UCD \\
2504 & Electronic, Optical and Magnetic Materials & TCD, UCD, Tyndall National Institute \\
2505 & Materials Chemistry & TCD, UL, UCD \\
2506 & Metals and Alloys & TCD, UCD, DCU \\
2507 & Polymers and Plastics & UL, UCD, UG \\
2508 & Surfaces, Coatings and Films & TCD, DCU, UCD \\
\multicolumn{3}{l}{\textbf{Mathematics}}\\
2600 & General Mathematics & UL, UCD, UG \\
2601 & Mathematics (miscellaneous) & UCD, DCU, TUD \\
2602 & Algebra and Number Theory & UCD, UG, TUD \\
2603 & Analysis & UCD, UG, R\&D Department, EI Electronics \\
2604 & Applied Mathematics & UL, UCD, UG \\
2605 & Computational Mathematics & UCD, UL, ATU \\
2606 & Control and Optimization & UG, UCD, UL \\
2607 & Discrete Mathematics and Combinatorics & UCD, DCU, MU \\
2608 & Geometry and Topology & UG, UCD, MTU \\
2609 & Logic & TCD, UCD, UL \\
2610 & Mathematical Physics & UCC, UCD, UG \\
2611 & Modeling and Simulation & TCD, UG, UCD \\
2612 & Numerical Analysis & UCD, UL, TUD \\
2613 & Statistics and Probability & MU, UL, TCD \\
2614 & Theoretical Computer Science & UCD, UG, TCD \\
\multicolumn{3}{l}{\textbf{Medicine}}\\
2700 & General Medicine & UCD, TCD, RCSI \\
2701 & Medicine (miscellaneous) & UCC, UCD, TCD \\
2702 & Anatomy & UCC, Cork University Hospital, DCU \\
2703 & Anesthesiology and Pain Medicine & University Hospital Galway, UCD, RCSI \\
2704 & Biochemistry (medical) & TCD, UCD, RCSI \\
2705 & Cardiology and Cardiovascular Medicine & UG, RCSI, University Hospital Galway \\
2706 & Critical Care and Intensive Care Medicine & St James's Hospital, TCD, RCSI \\
2707 & Complementary and alternative medicine & UL, UCD, National Rehabilitation Hospital \\
\end{tabular*}
\end{table*}

\begin{table*} 
\centering
\begin{tabular*}{\textwidth}{@{\extracolsep{\fill}} lll}
\hline
\hline
ASJC & Subject Area & Top Institutions \\
\hline 
2708 & Dermatology & UG, UCD, St Vincent's University Hospital \\
2709 & Drug Guides & TCD, Children's Health Ireland, St Vincent's University Hospital \\
2710 & Embryology & UCC, UL, UCD \\
2711 & Emergency Medicine & UCD, RCSI, Mater Misericordiae University Hospital \\
2712 & Endocrinology, Diabetes and Metabolism & UCD, TCD, UG \\
2713 & Epidemiology & UG, Health Research Board, UCC \\
2714 & Family Practice & RCSI, UL, SETU\\
2715 & Gastroenterology & UCC, RCSI, Cork University Hospital \\
2716 & Genetics (clinical) & Mater Misericordiae University Hospital, UG, St Vincent's University Hospital \\
2717 & Geriatrics and Gerontology & TCD, UCC, UCD \\
2718 & Health Informatics & UCD, TCD, UCC \\
2719 & Health Policy & UCD, TCD, RCSI \\
2720 & Hematology & RCSI, UG, St James's Hospital \\
2721 & Hepatology & UCC, TCD, St James's Hospital \\
2722 & Histology & UCC, ATU, UG \\
2723 & Immunology and Allergy & TCD, UCC, UCD \\
2724 & Internal Medicine & UCD, TCD, RCSI \\
2725 & Infectious Diseases & UCC, UCD, TCD \\
2726 & Microbiology (medical) & UCC, UCD, Teagasc \\
2727 & Nephrology & TCD, Beaumont University Hospital, UCD \\
2728 & Neurology (clinical) & TCD, UG, Health Research Board \\
2729 & Obstetrics and Gynecology & UCD, National Maternity Hospital, UCC \\
2730 & Oncology & TCD, UCD, St Vincent's University Hospital \\
2731 & Ophthalmology & TUD, UCD, Retina International \\
2732 & Orthopedics and Sports Medicine & UCD, Sports Surgery Clinic, UL \\
2733 & Otorhinolaryngology & RCSI, TCD, Victoria University Hospital \\
2734 & Pathology and Forensic Medicine & UG, UCC, Cork University Hospital \\
2735 & Pediatrics, Perinatology and Child Health & UCC, Children's Health Ireland, RCSI \\
2736 & Pharmacology (medical) & UG, RCSI, UCD \\
2737 & Physiology (medical) & TCD, DCU, UCD \\
2738 & Psychiatry and Mental health & TCD, UCD, MU \\
2739 & Public Health, Environmental and Occupational Health & TCD, UG, UCD \\
2740 & Pulmonary and Respiratory Medicine & RCSI, UCD, St James's Hospital \\
2741 & Radiology, Nuclear Medicine and imaging & UCC, UCD, UG \\
2742 & Rehabilitation & UCD, UCC, UL \\
2743 & Reproductive Medicine & UG, UCD, UCC \\
2744 & Reviews and References (medical) & UL, UCC, UCD \\
2745 & Rheumatology & TCD, UCD, St Vincent's University Hospital \\
2746 & Surgery & RCSI, Mater Misericordiae University Hospital, UG \\
2747 & Transplantation & Cork University Hospital, Mater Misericordiae University Hospital, UL \\
2748 & Urology & Mercy University Hospital, UCC, TCD \\
\multicolumn{3}{l}{\textbf{Neuroscience}}\\
2800 & General Neuroscience & TCD, TUD, UCC \\
2801 & Neuroscience (miscellaneous) & TCD, UCC, TUD \\
2802 & Behavioral Neuroscience & UCC, RCSI, TCD \\
2803 & Biological Psychiatry & TCD, UG, UCC \\
2804 & Cellular and Molecular Neuroscience & UCC, TCD, UCD \\
2805 & Cognitive Neuroscience & TCD, UG, UCD \\
2806 & Developmental Neuroscience & RCSI, TCD, UG \\
2807 & Endocrine and Autonomic Systems & RCSI, UCC, TCD \\
2808 & Neurology & TCD, UCC, UG \\
2809 & Sensory Systems & UCD, Royal Victoria Eye and Ear Hospital \\
\multicolumn{3}{l}{\textbf{Nursing}}\\
2900 & General Nursing & UCD, UL, UCC \\
2901 & Nursing (miscellaneous) & RCSI, UCC, SETU\\
2902 & Advanced and Specialized Nursing & TCD, Health Research Board, UCD \\
2903 & Assessment and Diagnosis & UCD, UCC, RCSI \\
2904 & Care Planning & SETU, TCD, Trinity St James's Cancer Institute \\
2905 & Community and Home Care & Health 4 U, SETU, UG \\
2906 & Critical Care Nursing & RCSI, UCD, University Hospital Galway \\
2907 & Emergency Nursing & UCD, UL, University Hospital Limerick \\
2908 & Fundamentals and skills & UCD, SETU, RCSI \\
2909 & Gerontology & RCSI, UCC, Cork University Hospital \\
2910 & Issues, ethics and legal aspects & UCD, UG, DCU \\
2911 & Leadership and Management & DCU, UCD, RCSI \\
2912 & LPN and LVN & TCD, UCD, UL \\
\end{tabular*}
\end{table*}

\begin{table*} 
\centering
\begin{tabular*}{\textwidth}{@{\extracolsep{\fill}} lll}
\hline
\hline
ASJC & Subject Area & Top Institutions \\
\hline
2913 & Maternity and Midwifery & UL, UCC, UCD \\
2914 & Medical–Surgical Nursing & UG, UL, UCC \\
2915 & Nurse Assisting & RCSI \\
2916 & Nutrition and Dietetics & UCC, UCD, TCD \\
2917 & Oncology (nursing) & UG, TCD, RCSI \\
2918 & Pathophysiology & UCC, Cork University Hospital \\
2919 & Pediatrics & TCD, RCSI, UL \\
2921 & Psychiatric Mental Health & UCC, TCD, UG \\
2922 & Research and Theory & RCSI, UCD, UCC \\
2923 & Review and Exam Preparation & UG, RCSI, UCC \\
\multicolumn{3}{l}{\textbf{Pharmacology, Toxicology and Pharmaceutics}}\\
3000 & General Pharmacology, Toxicology and Pharmaceutics & RCSI, UG, ATU \\
3001 & Pharmacology, Toxicology and Pharmaceutics (miscellaneous) & UCD, RCSI, TUD \\
3002 & Drug Discovery & Teagasc, UCC, UCD \\
3003 & Pharmaceutical Science & UCC, SETU, UL \\
3004 & Pharmacology & UCC, RCSI, UCD \\
3005 & Toxicology & UG, UCD, RCSI \\
\multicolumn{3}{l}{\textbf{Physics and Astronomy}}\\
3100 & General Physics and Astronomy & TCD, UCD, UG \\
3101 & Physics and Astronomy (miscellaneous) & ATU, UCD, TCD \\
3102 & Acoustics and Ultrasonics & TCD, UCD, Nokia Bell Labs \\
3103 & Astronomy and Astrophysics & TCD, DIAS, ATU \\
3104 & Condensed Matter Physics & UCD, TCD, UG \\
3105 & Instrumentation & UCD, UG, UCC \\
3106 & Nuclear and High Energy Physics & TCD, UCD, ATU \\
3107 & Atomic and Molecular Physics, and Optics & TCD, UCD, TUD \\
3108 & Radiation & UCD, UCC \\
3109 & Statistical and Nonlinear Physics & TCD, UCD, MU \\
3110 & Surfaces and Interfaces & DCU, TCD, UL \\
\multicolumn{3}{l}{\textbf{Psychology}}\\
3200 & General Psychology & UCD, UCC, DCU \\
3201 & Psychology (miscellaneous) & MU, TCD, UCD \\
3202 & Applied Psychology & UL, TCD, DCU \\
3203 & Clinical Psychology & TCD, MU, UCD \\
3204 & Developmental and Educational Psychology & DCU, UCC, UG \\
3205 & Experimental and Cognitive Psychology & UL, TCD, UCD \\
3206 & Neuropsychology and Physiological Psychology & TCD, UG, UL \\
3207 & Social Psychology & UCD, UL, UCC \\
\multicolumn{3}{l}{\textbf{Social Sciences}}\\
3300 & General Social Sciences & UCD, MU, UCC \\
3301 & Social Sciences (miscellaneous) & DCU, UCD, TCD \\
3302 & Archeology & UCD, UCC, MU \\
3303 & Development & UCD, DCU, TCD \\
3304 & Education & DCU, UL, UCD \\
3305 & Geography, Planning and Development & UCD, UG, DCU \\
3306 & Health (social science) & UCD, UCC, TCD \\
3307 & Human Factors and Ergonomics & UCD, UL, TUD \\
3308 & Law & DCU, MU, TCD \\
3309 & Library and Information Sciences & UG, TCD, UCD \\
3310 & Linguistics and Language & DCU, UCD, UG \\
3311 & Safety Research & DCU, UCD, TUD \\
3312 & Sociology and Political Science & UCD, DCU, TCD \\
3313 & Transportation & TCD, DCU, UCD \\
3314 & Anthropology & UCD, TCD, UG \\
3315 & Communication & DCU, UCD, UL \\
3316 & Cultural Studies & DCU, UCD, MU \\
3317 & Demography & UCD, TCD, MU \\
3318 & Gender Studies & DCU, MU, TCD \\
3319 & Life-span and Life-course Studies & TCD, UCC, DCU \\
3320 & Political Science and International Relations & UCD, DCU, UCC \\
3321 & Public Administration & TCD, DCU, UCD \\
3322 & Urban Studies & UCD, TCD, MU \\
\end{tabular*}
\end{table*}

\begin{table*} 
\centering
\begin{tabular*}{\textwidth}{@{\extracolsep{\fill}} lll}
\hline
\hline
ASJC & Subject Area & Top Institutions \\
\hline
\multicolumn{3}{l}{\textbf{Veterinary}}\\
3400 & General Veterinary & UCD, Teagasc, UL \\
3401 & Veterinary (miscellaneous) & UCC, TCD, Teagasc \\
3402 & Equine & UCD, UL \\
3403 & Food Animals & Teagasc, Cork Pet Behaviour Centre, Beaumont University Hospital \\
3404 & Small Animals & UCD, UL \\
\multicolumn{3}{l}{\textbf{Dentistry}}\\
3500 & General Dentistry & TCD, Dublin Dental University Hospital, RCSI \\
3501 & Dentistry (miscellaneous) & Dublin Dental University Hospital, TCD \\
3502 & Dental Assisting & Dublin Dental University Hospital, TCD \\
3503 & Dental Hygiene & SETU, Dublin Dental University Hospital, TCD \\
3504 & Oral Surgery & TCD, UCC, Dublin Dental University Hospital \\
3505 & Orthodontics & Dublin Dental University Hospital, TCD \\
3506 & Periodontics & Dublin Dental University Hospital, TCD \\
\multicolumn{3}{l}{\textbf{Health Professions}}\\
3600 & General Health Professions & UL, MU, DCU \\
3601 & Health Professions (miscellaneous) & UCC, UCD, Teagasc \\
3602 & Chiropractics & UCC, TCD, Trinity St James's Cancer Institute \\
3603 & Complementary and Manual Therapy & UL, UG, SETU\\
3604 & Emergency Medical Services & UL, UCC, Letterkenny University Hospital \\
3605 & Health Information Management & UCD, UCC, TCD \\
3606 & Medical Assisting and Transcription & RCSI \\
3607 & Medical Laboratory Technology & TCD, UCD, UG \\
3608 & Medical Terminology & Ekrity, Vistatec \\
3609 & Occupational Therapy & UCC, TCD, UG \\
3610 & Optometry & TUD, Children's Health Ireland \\
3611 & Pharmacy & UCC, RCSI, TCD \\
3612 & Physical Therapy, Sports Therapy and Rehabilitation & UCD, DCU, UL \\
3613 & Podiatry & Cork University Hospital, University Hospital Galway, RCSI \\
3614 & Radiological and Ultrasound Technology & UCC, UCD, UG \\
3615 & Respiratory Care & Glenville Nutrition Ireland \\
3616 & Speech and Hearing & TCD, UL, UG \\
\end{tabular*}
\end{table*}

\end{document}